\documentclass[useAMS,usenatbib]{mn2e}
\usepackage{psfig,ifthen,dcolumn,url}

\newcommand{\Ttau} {\hbox{$T(\tau)$}-relation}
\newcommand{\stimes}{\mbox{$\times$}}
\newcommand{\nsp}{\!\!\!\!}

\newcommand{\ion}[2]{\mbox{#1\,{\sc #2}}}
\newcommand{\figdir}{figs}
\newcommand{\mfigur}[3]
    {\begin{figure}
        \centerline{\psfig{figure=\figdir/#1.ps,height=#2}}
        \caption{\label{#1}#3}
    \end{figure}}
\def\note #1]{{\bf #1]}}
\def\aatab#1#2#3#4#5#6{\ifthenelse{\equal{*}{#1}}
{\begin{table*}
\caption[]{\label{#2} #3}
  \begin{flushleft}
    \begin{tabular}{#4}
      \hline\noalign{\smallskip} #5
          \noalign{\smallskip} \hline \noalign{\smallskip} #6
          \noalign{\smallskip} \hline
        \end{tabular}
  \end{flushleft}
\end{table*}}
{\begin{table}
\caption[]{\label{#1} #2}
  \begin{flushleft}
    \begin{tabular}{#3}
      \hline\noalign{\smallskip} #4
      \noalign{\smallskip} \hline \noalign{\smallskip} #5
      \noalign{\smallskip} \hline
    \end{tabular}
  \end{flushleft}
\end{table}}}
\newcommand{\oh}[1]{\omit\hidewidth #1\hidewidth}
\newcommand{\be}[1]{\begin{equation}\label{#1}}
\newcommand{\ee}{\end{equation}}
\newcommand{\bea}[1]{\begin{eqnarray}\label{#1}}
\newcommand{\eea}{\end{eqnarray}}

\newcommand{\mfrac}[2]{{\textstyle{#1\over #2}}}
\newcommand{\sfrac}[2]{{\leavevmode
                    \kern.0em\raise.5ex\hbox{\the\scriptfont0 #1}
                    \kern-0.07em\raise.25ex\hbox{{\small{/}}}
                    \kern-0.07em\lower.00ex\hbox{\the\scriptfont0 #2}}}
\renewcommand{\d}{{\rm d}}
\newcommand{\dd} [2]{\frac{{\rm d}{#1}}{{\rm d}{#2}}}
\newcommand{\eg}  {e.g.}            
\newcommand{\cf}  {cf.}             
\newcommand{\ie}  {i.e.}            
\newcommand{\etc} {etc.}
\begin{document}
    \title[Improvements to Stellar Structure Models: $T(\tau)$-Relations]%
          {Improvements to Stellar Structure Models, Based on a Grid of 3D
                        Convection Simulations. I. $T(\tau)$-Relations}

    \author[R. Trampedach, et al.]{Regner Trampedach$^{1,2}$\thanks{E-mail: {\protect\url{trampeda@lcd.colorado.edu}}},
     Robert F. Stein$^{3}$,
     J{\o}rgen {Christensen-Dalsgaard}$^{1}$, \newauthor
     {\AA}ke Nordlund$^{4}$ and
     Martin Asplund$^{5}$\\
    $^{1}$Stellar Astrophysics Centre, Department of Physics and Astronomy,
          Aarhus University, DK--8000 Aarhus C, Denmark\\
    $^{2}$JILA, University of Colorado and National Institute of Standards
          and Technology, 440 UCB, Boulder, CO 80309, USA\\
    $^{3}$Department of Physics and Astronomy, Michigan State University
          East Lansing, MI 48824, USA\\
    $^{4}$Astronomical Observatory/Niels Bohr Institute, Juliane Maries Vej 30,
          DK--2100 Copenhagen {\O}, Denmark\\
    $^{5}$Research School of Astronomy and Astrophysics,
          Mt.\ Stromlo Observatory, Cotter Road, Weston ACT 2611, Australia}

    \date{Received \today / Accepted \today}
    \pagerange{\pageref{firstpage}--\pageref{lastpage}} \pubyear{2013}

    \maketitle

    \label{firstpage}

    \begin{abstract}
Relations between temperature, $T$, and optical depth, $\tau$,
are often used for describing the photospheric transition
from optically thick to optically thin in stellar structure models.
We show that this is well justified, but also that currently used
{\Ttau}s are often inconsistent with their implementation.
As an outer boundary condition on the system of stellar structure
equations, {\Ttau}s have an undue effect on the overall structure of stars.
In this age of precision asteroseismology, we need to re-assess both
the method for computing and for implementing {\Ttau}s, and the
assumptions they rest on.

We develop a formulation for proper and consistent evaluation of
{\Ttau}s from arbitrary 1D or 3D stellar atmospheres, and for their
implementation in stellar structure and evolution models.

We extract radiative {\Ttau}s, as described by our new formulation,
from 3D simulations of convection in deep stellar atmospheres
of late-type stars from dwarfs to giants.
These simulations employ realistic opacities and equation of state,
and account for line-blanketing.
For comparison, we also extract {\Ttau}s from 1D MARCS model
atmospheres using the same formulation.

$T(\tau)$-relations from our grid of 3D convection simulations
display a larger range of behaviours with surface gravity, compared
with those of conventional theoretical 1D hydrostatic atmosphere models
based on the mixing-length theory for convection. The 1D atmospheres show
little dependence on gravity. 1D atmospheres of main-sequence stars also
show an abrupt transition to the diffusion approximation at $\tau\simeq 2.5$,
whereas the 3D simulations exhibit smooth transitions that occur at the same
depth for $M\simeq 0.8$\,M$_\odot$, and higher in the atmosphere for both
more and less massive main-sequence stars.
Based on these results, we recommend no longer to use scaled solar {\Ttau}s.

Files with {\Ttau}s
for our grid of simulations are made available to the community,
together with routines for interpolating in this irregular grid. We also
provide matching tables of atmospheric opacity, for consistent implementation
in stellar structure models.
    \end{abstract}


\section{Introduction}
\label{intro}

The complexity of stellar atmospheres is due to the highly complicated
behaviour of opacity with wavelength and thermodynamic state,
coupled with the non-local nature of radiative transfer in the photospheric
transition from optically thick to optically thin.
To treat this region properly, present-day stellar atmosphere codes solve the
radiative transfer for
hundreds of thousands of wavelength points, including tens of millions of
spectral lines.
Further ``non-classical'' complications arise when deviations from
local thermodynamic equilibrium (LTE) need to be accounted for, when the
absence of convection allows for chemical stratification, or when 
strong magnetic fields or rapid rotation affect the atmospheric structure.
Thus, although stellar evolution modelling requires
non-trivial boundary conditions at the stellar surface,
these complications obviously render impractical
the direct calculation of realistic atmospheres in stellar structure models.

A solution to this problem is to use precomputed results of stellar atmosphere modelling
(semi-empirical or fully theoretical) as upper boundary conditions
for stellar structure models.
Knowing pressure (equation-of-state, EOS) and Rosseland opacity as functions of $\varrho$ and $T$, and assuming
hydrostatic equilibrium, the system of equations can be closed by the {\Ttau}
without having to solve the frequency-dependent radiative transfer --
{\ie}, the stratification of the detailed atmosphere calculation can be
recovered with a mean opacity, as we show in Sect.~\ref{Ttaubasis}.
Note that the Rosseland opacity is \emph{not} used for performing radiative
transfer in the atmosphere of the structure model---it is merely used to
reconstitute the detailed atmosphere model, which has been condensed into the
{\Ttau}. The resulting atmosphere is therefore not grey, unless the 
 grey {\Ttau} has been used (see, e.g., Fig. \ref{ttau_scan}).

Since this method was implemented by, {\eg}, \citet{boehm:mlt}, this outer
boundary has received relatively little attention. Indeed, {\Ttau}s from that period are
still in widespread use in stellar structure models -- despite these {\Ttau}s
having little resemblance with current atmosphere models.

\citet{chabrier:LowM-evol} investigated the effect of {\Ttau}s on
stellar evolution for solar-composition low-mass stars. The authors were
rightly concerned about the
proper implementation and interpretation of {\Ttau}s in the transition
between radiative and convective zones -- an issue which is often overlooked.
\citet{vandenberg:T-tau} performed a similar analysis of stellar models using
MARCS stellar atmosphere models \citep{MARCS-2} as outer boundaries, as well
as the often used grey and
\citet[see Sect.\ \ref{solarT-tau}]{krishna-swamy:lines-K-dwarfs} atmospheres.
They were likewise concerned about how and where to merge the two formulations,
and the inconsistencies and discontinuities arising there.
The fundamental problem is that the transition should ideally take place in
the optically deep layers, where the diffusion approximation applies, and above the
top of the convection zone to avoid inconsistencies due to differing formulations of
convection. It is often assumed that optically deep means $\tau\ga 1$ in
which case it is barely possible to have this separation in one dimensional (1D) models.
\citet{morel:T-tau-starmods} show, however, that the diffusion approximation
is not fulfilled unless $\tau\ga 10$ (as we confirm).
Furthermore, in more realistic 3D simulations of convective atmospheres, even
$\tau = 1$ overlaps with convection. The problems plaguing the outer boundary
conditions are recognised, yet there has been little theoretical progress on
the issue.
We address these problems with our new self-consistent formulation, discussed in
Sect.~\ref{Ttaubasis}, which also eliminates the extra parameters introduced by
the above mentioned prescriptions.

A helioseismic analysis of various commonly used prescriptions for the
outer boundary of solar models was carried out by \citet{morel:T-tau-starmods}.
They found sizeable effects that show up as part of the so-called \emph{surface
effect}: a systematic and frequency dependent shift of frequencies
\citep{jcd:solar-freq-shifts}. They also show how the classic calibration of
solar models to the present Sun \citep{gough:MLT-calibr}, tightly couples the
outer boundary condition to the mixing-length of convection.

Since the solar {\Ttau} can in principle be inferred
from limb-darkening observations \citep[e.g.,][]{mitchell:LT-Ttau},
the use of semi-empirical models based on
such observations has often been considered the safest choice. The observations
and corresponding {\Ttau}s can only be performed monochromatically, however,
which is of little use to stellar modellers. Connecting to the more useful
Rosseland optical depth scale, requires a pan-chromatic knowledge of
opacities, which is the exact weakness of fully theoretical models that
semi-empirical models seek to circumvent. At the time there were indications,
though, that the Rosseland opacity was little affected by spectral lines, in
which case the 5000\,{\AA} monochromatic opacity (which is straightforward to compute, since only few and well-known sources contribute to that opacity)
was a good proxy in late-type
stars. This justified the use of semi-empirical models, 
as a largely model-independent alternative to fully theoretical models.
Subsequent work has shown
that spectral lines can have a large influence on the Rosseland opacity,
adding 20--40\% to the opacity in stellar atmospheres
(see Trampedach, in prep.).
Since the Rosseland mean is a harmonic mean, it heavily weights the lowest
opacity, that is, the continuum. When the spectral density of lines becomes
so high that lines overlap, a pseudo-continuum is formed, which raises the
Rosseland mean \citep[][for analysis of interior opacities]{rog-igl:science}.
The result is that semi-empirical models
depend almost as much on opacities (and other atomic physics), as do the fully
theoretical solar model atmosphere.

Theoretical {\Ttau}s from 1D stellar atmosphere models have been published
in connection with, {\eg}, the {\sc ATLAS} \citep{kur:modatm,kur:atlas12,kurucz:newATLAS9},
the MARCS \citep{gus:modgrid,asplund:RCorBor,MARCS-2},
the NextGen \citep{hauschildt:nextgen-atm,hauschildt:nextgen-grid2,short:NLTE-Phoenix}
and the MAFAGS-OS \citep{grupp:MAFAGS-OS-1,grupp:MAFAGS-OS-NewAtomic}
grids of stellar atmospheres.
These are grids in effective temperature, surface gravity and metallicity
and the grids are dense enough that simple interpolation is safe. The level of
sophistication is very impressive, but detailed comparisons with solar spectra
still reveal many lines unaccounted for \citep[e.g.,][]{kurucz:InclAllLines}.
\citet{plez:1DatmsForGaia} compare the first three grids for late-type stars
and find good agreement between their atmospheric structures
\citep[see also][]{MARCS-2}, but significant differences between their spectra.

In late-type stars, however, the
treatment of convection is the weakest point in the modelling of atmospheres.
Not only are they not one-dimensional, but
the convective fluctuations are also well outside the regime of linear perturbations.
So far, the only way to deal with the combined problem of radiation and
convection in stellar photospheres is to perform realistic radiation
hydrodynamic (RHD) simulations. There are now a couple of grids of such
simulations in development or available: the CIFIST grid
\citep{ludwig:CIFIST-grid} of CO$^5$BOLD simulations
\citep{CO5BOLD,freytag:CO5BOLD} and the Stagger Grid
\citep{remo:StaggerGrid,magic:stagger-grid} of Stagger-code simulations
\citep{nordlund:StaggerCode,bob:SuperGran}.
{\Ttau}s were briefly discussed by \citet{ludwig:alfa-cal} for an
earlier grid, using a 2D version of the predecessor of the CO$^5$BOLD code.
They used mostly grey radiative transfer in the simulations and a fit to the
grey atmosphere in the structure models, with no details on their
implementation of the latter. The few {\Ttau}s they did
evaluate from non-grey simulations were not published.
\citet{magic:stagger-alpha} used 1D atmosphere models corresponding to the
simulations of their stagger-grid, as their {\Ttau}s, also not published.
Tanner et al. (2014) carried out 3D simulations that use the Eddington
approximation to the grey atmosphere, instead of solving the radiative transfer,
and explored effects of metallicity. This approach neatly isolates the effect
of adiabatic cooling by convective overshooting
\citep{nordlund+stein:RadDyn1991,asplund:low-Z-Li,remo:3DRedGiantAbunds},
since they ignore line-blanketing \citep{chandrasekhar:LineBlanket} which would
normally interfere with this effect. It does, however, not describe a realistic
atmosphere, since the defining radiative transfer has been ignored.

We base our present work on the
grid of solar-metallicity, deep convective atmospheres by
\citet{trampedach:3Datmgrid}, which used the \citet{stein:solar-granI}-code.
We describe the most pertinent features of these convection simulations in
Sect.~\ref{sims}
and derive a consistent formulation of {\Ttau}s that also works with 3D
radiative transfer
in Sect.~\ref{av-proc}. The {\Ttau}s
of the simulations are presented in Sect.~\ref{simT-tau} and
in Sect.~\ref{solarT-tau}, we make a more detailed analysis of our
solar simulation and compare with semi-empirical and theoretical {\Ttau}s
from the literature.
The general variation of our {\Ttau}s with stellar parameters, is explored in
Sect.~\ref{parvar}, by means of interpolation routines readily employed
in stellar structure calculations.

This paper is the first in a series dedicated to the improvement of stellar
structure and evolution calculations. These improvements are based on lessons
learnt from 3D radiation-coupled hydrodynamical simulations of convection in
the atmospheres of late-type stars (F-K). Here we present
results on the radiative part of the mean stratification of the simulations
in the form of {\Ttau}s, to be used as outer boundary conditions for stellar
structure models.
Paper II \citep{trampedach:alfa-fit} deals with the convective part of the
mean stratification by calibrating the mixing length and presenting the results
in a form easy to implement in stellar structure codes.
The radiative and the convective parts of the problem are strongly
interdependent, as discussed in the present paper and in paper II,
but we also show that
separating the two parts is possible and that such a separation is relevant for 
stellar structure models.
Paper\,II also examines the coupling between {\Ttau}s and the mixing-length
of convection.
A future paper
will address the consequences of applying
the above improvements to stellar evolution calculations.

\section{The Basis for $T(\tau)$-Relations}
\label{Ttaubasis}

A {\Ttau} is needed for describing the photospheric transition between
optically thick (diffusion approximation) and optically thin (the free-streaming
approximation) layers. In late-type stars this transition is affected by the large
temperature fluctuations caused by convection, but it is in principle separable from the
issue of how large a fraction of the flux is transported by convection.
We first need to formalise these statements and
establish the theoretical foundation for a consistent definition
and use of {\Ttau}s. 

The 3D simulations are all performed with uniform gravity, which corresponds to
the plane-parallel approximation in 1D. We keep the following derivations in
that approximation.
This should be valid even for our lowest gravity simulations
(Nos.\ 1 and 2 in Table \ref{simlist}, with $\log g=2.2$),
since \citet{plez-aake:OS-Mgiants} found a maximum sphericity effect
of a mere $-20$\,K for a 3\,800\,K, $\log g=1.0$, 1\,$M_\odot$ giant, at $\log\tau=-4.5$, using MARCS 1D model atmospheres.
The sphericity effect will obviously be smaller for larger $g$ as in our case.

We describe the 1D radiative transfer in terms of the usual
moments of the radiation field
\be{moments}
    I^{(n)}_\lambda=\mfrac{1}{2}\int_{-1}^1 \mu^nI_\lambda(\mu){\rm d}\mu\ ,
\ee
where the intensity, $I_\lambda$, only depends on the angle with the surface
normal, $\mu=\cos\theta$. Dependence on optical depth, $\tau$, is implied
throughout this section. Extension to the 3D case is dealt with 
in Sect.~\ref{av-proc}. The negative half of the integral accounts for photons
going into the interior. At the top of the domain (if $\tau\ll 1$), this can
be assumed zero, unless irradiation by a companion star has to be accounted for.

The first three moments are also called
\be{JHK}
    J_\lambda=I^{(0)}_\lambda\,,\quad \frac{F_{\rm rad,\lambda}}{4\pi} =
        H_\lambda=I^{(1)}_\lambda{\rm~~and}\quad
        K_\lambda=I^{(2)}_\lambda\ ,
\ee
where $F_{\rm rad,\lambda}$ is the monochromatic astrophysical flux.
The transfer equation is
\be{transfer}
    \mu\dd{I_\lambda(\mu)}{\tau_\lambda} = I_\lambda(\mu) - S_\lambda\ ,
\ee
where the source-function, $S_\lambda$, is assumed to be isotropic.

The corresponding radiative heating (cooling when negative) is
\be{qrad}
    Q_{\rm rad,\lambda} = 4\pi\varrho\kappa_\lambda(J_\lambda-S_\lambda)\ ,
\ee
where the $4\pi$ comes from the angular integration of an isotropic quantity.
$Q_{\rm rad,\lambda}$ is the extensive (per unit volume) radiative heating.
Since any heating mechanism, $X$, has an associated flux $F_X$, given by
$Q_X = {\rm d}F_X/{\rm d}z$,
$Q_X$ is also known as the flux divergence.

A solution in radiative equilibrium obviously obeys
$Q_{\rm rad}=\int_0^\infty Q_{\rm rad,\lambda}{\rm d}\lambda=0$.
In the more general case where convection also supplies heating or cooling,
but no energy sinks or sources are present,
the equilibrium constraint is $Q_{\rm rad} + Q_{\rm conv} = 0$. How this convective term affects
the equilibrium stratification can be gleaned from the angular moments of the
transfer equation, Eq.~(\ref{transfer}). The zeroth moment
\be{rad0}
    \dd{H_\lambda}{\tau_\lambda} =
    \frac{1}{4\pi}\dd{F_{\rm rad,\lambda}}{\tau_\lambda} = J_\lambda - S_\lambda\ ,
\ee
contains the derivative of $H_\lambda$ and the first moment
\be{rad1}
    \dd{K_\lambda}{\tau_\lambda} = H_\lambda = \frac{F_{\rm rad,\lambda}}{4\pi}\ .
\ee
contains $H_\lambda$ itself.
Since we have no energy generated in the atmospheres, the luminosity is
constant throughout the atmosphere translating into a constant flux 
$F_{\rm tot} = 4\pi H + F_{\rm conv}$ 
in the plane-parallel case,
where $H$ is the wavelength-integrated $H_\lambda$.
The decrease of $H$ as we enter the convection zone,
provides the first-order effect of convection on the $T(\tau)$ stratification.

\subsection{Convective vs. Radiative $T(\tau)$-Relations}
\label{Tradtau}

The 
temperature as a function of optical depth, $T(\tau)$,
is trivial to extract from any atmosphere
model, be it 1D or a 3D simulation.
The {\Ttau}, however, is greatly affected by the presence of convection,
having a much smaller gradient in convective regions (see Fig.~\ref{ttau_scan}).
When applying
{\Ttau}s to stellar structure calculations, the convective fluxes will most
likely differ between the atmosphere model and the 1D structure model, which
renders the {\Ttau} inappropriate for the structure model. Using the {\Ttau}
from even the most sophisticated 3D simulation of a convective atmosphere in a
1D model would be both inconsistent and unphysical if the convective 
flux differs between the two cases.
Unfortunately we do not yet have a way of
directly incorporating the convection of the 3D simulations into the 1D models
in any consistent or even physically meaningful way. To rectify this is
obviously a long-term goal of ours.

The first-order effect of convection on {\Ttau}s is due to the radiative flux
not being constant. If we can somehow express the {\Ttau} with an explicit
appearance of the radiative flux, it should be possible to isolate this
first-order effect and calculate a \emph{radiative} {\Ttau} ({\ie}, reduced
to the radiative equilibrium case with
$Q_{\rm rad}=0$ and therefore $F_{\rm rad}={\rm const}$). When used in stellar
structure calculations the first-order convective effect can be added back in,
as appropriate for that particular model (see sect.\ \ref{Ttau-impl}). From
here on we will call temperatures from {\Ttau}s with this first-order
convective effect subtracted, $T_{\rm rad}(\tau)$---the radiative
{\Ttau}.

The original, unaltered {\Ttau} of the atmosphere models, which includes the
transition to convective transport of the flux, will be referred to as
partially convective {\Ttau}s.

Another effect of a varying $H$, is the finite difference between source
function and mean intensity, $\Delta=J-S$, giving rise to the radiative
heating or flux divergence of Eq.~(\ref{qrad}). In radiative equilibrium
$\Delta$ converges, $\Delta\rightarrow 0$, with optical depth, but not in the transition to
a convection zone. The relative difference, $\Delta/S$, is small, however. The
largest values are found in the cooling peak at the top of convective
envelopes, where $|\Delta|/S\la 1\stimes 10^{-3}$, or in a heating bump above
the photosphere of the coolest dwarf stars, with $|\Delta|/S\la 5\stimes
10^{-3}$. Estimating the effect on temperature as $|\Delta|/S\simeq 4\Delta T$,
we get $|\Delta T|\la 1$\,K and $|\Delta T|\la 4$\,K, respectively, and we have
therefore chosen to ignore this effect. Since it is equally ignored in the
1D models used here, this in practice means that we assume the same $\Delta$
in the atmosphere model as in the 1D structure model. The actual error
thus committed is therefore much smaller.

It follows that in order to make self-consistent 1D stellar models
we need to remove the first-order effect from the 3D {\Ttau}: the transition
from convective to radiative transport of energy in the photosphere.
All higher-order convective effects arising from, {\eg}, the
mean stratification differing from the 1D model, cooling by convective
overshooting (cf.~Sect.~\ref{parvar}), the large convective
fluctuations in the photosphere and the correlations between them will,
however, be retained in $T_{\rm rad}(\tau)$. And this is plenty motivation for
continuing this exercise, as will be shown in the comparisons between
1D and 3D {\Ttau}s in Sect.~\ref{simT-tau}.

\subsection{$T(\tau)$-Relations from Grey Atmospheres}
\label{grey}

In order to isolate the first-order convective effect on the {\Ttau} we seek
an expression containing $H$, and to simplify the problem we begin with the
grey case (by simply dropping the $\lambda$ subscripts), and generalise
afterwards.

The choice of formulation rests on the choice of quantity that will be assumed
invariant under convection, {\ie}, unchanged whether convection carries
flux or not. The invariant should contain quantities from Eqs.~(\ref{rad0})
and (\ref{rad1}) which can be used to reconstruct the Planck function,
$B=\sigma T^4/\pi$.
Reasonable choices are $K/B$, $K/J$, d$K/$d$B$ or d$K/$d$J$, which all can be
shown to converge to 1/3 for $\tau \rightarrow \infty$
\citep{mihalas:stel-atm,rutten:RadTransf}. These ratios can all 
be used in a similar way as the {\em variable Eddington factor}, $K/J$
\citep{auer:VarEddingtonFacts}.
We choose
$f_{{\rm d}B}(\tau) \equiv {\rm d}K/$d$B$ for the relative simplicity of the
temperature reconstruction. We effectively distil the whole stellar atmosphere
calculation down to one quantity: $f_{{\rm d}B}$ as function of depth.
This is the essence of the elaborate atmosphere calculation that we want to
transfer to the stellar structure calculation, regardless of how convection
differs between the two cases.

Such a temperature perturbation, from different treatments of convection, would
cause differences in $f_{{\rm d}B}$ if it were to be computed anew from
radiative transfer on the perturbed structure. So by keeping $f_{{\rm d}B}$
invariant we impose on the structure model the radiative transfer result from
the full atmosphere calculation based on the latter's temperature structure. 
This counts amongst the second-order effects of convection on the radiative
transfer (variation of $f_{{\rm d}B}$ is indeed quadratic in small temperature
perturbations, but also sizeable at relevant amplitudes).

The main thrust in this paper is the utility of 3D atmosphere simulations
in setting outer boundary conditions for stellar structure models. Our method
can also be used with 1D stellar atmospheres, however, e.g., using different
formulations or parameters for convection than the interior models, but with a
seamless combination of the two (see Sect.\ \ref{Ttau-impl}).

Using Eq.~(\ref{rad1}) to express our invariant as
\be{fdB}
    f_{{\rm d}B}(\tau) \equiv \dd{K}{B}
    = \frac{H}{{\rm d}B/{\rm d}\tau} \rightarrow \frac{1}{3}
    \qquad{\rm for}\quad \tau \rightarrow \infty\ ,
\ee
it is straightforward to isolate the derivative of $B$ and integrate to
obtain
\be{Bint}
    B_x(\tau)=B(\tau_0)+\int_{\tau_0}^\tau\frac{H_x(\tau')}
                                    {f_{{\rm d}B}(\tau')}{\rm d}\tau'\ ,
\ee
where $H_x$ is the radiative flux of the stellar structure model, $x$, seeking
to recover the {\Ttau} of the atmosphere model. In other words, $x$ is the
structure model employing the {\Ttau} of the detailed atmosphere model.
$B_x$ is the corresponding Planck function.
The important point to realise here is that $H$ in Eq.~(\ref{fdB}) is that
of the detailed atmosphere calculation, from which the {\Ttau} is derived.

We transform to the actual {\Ttau} through the so-called
\emph{Hopf function}, $q(\tau)$, originally introduced to describe the solution
of grey radiative transfer in a semi-infinite atmosphere
\citep{hopf:HopfFunc,king:GrayAtm}. Here we use the same
formulation for the general case, and reserve $q_{\rm grey}$ for the original
intent.
The (generalised) Hopf function, $q(\tau)$, is then defined as
\be{Hopf}
    \frac{4}{3}\left(\frac{T}{T_{\rm eff}}\right)^4
                            = q(\tau) + \tau\ ,
\ee
on some optical depth scale, $\tau$.

The radiative Hopf function
(which assumes an atmosphere in radiative equilibrium)
is convergent,
$q_{\rm rad}(\tau)\rightarrow q_\infty$ for $\tau\rightarrow\infty$, and thus recovers the
diffusion approximation at depth.
Equivalently we describe a (partly) convective stratification, $T(\tau)$, by
$q_{\rm conv}(\tau)$. With part of the flux transported by convection, the
temperature is decoupled from the optical depth and there is no convergence
with depth. The Schwarzschild criterion \citep[e.g.,][]{SSE} for convective
instability tells us that at any depth the energy transport mechanism with the
smallest temperature gradient is the relevant one. The temperature in a partly
convective stratification will therefore be lower than in a corresponding
radiative one, and from Eq.\ (\ref{Hopf}) we see that $q_{\rm conv}(\tau)$ will
diverge to large negative values.

From Eq.\ (\ref{Bint}) we find that the radiative Hopf function,
$q_{\rm rad}(\tau)$, of a stratification in radiative equilibrium,
$T_{\rm rad}$, related by Eq.\ (\ref{Hopf}) is
\bea{qqrad}
    q_{\rm rad}(\tau)
    &=& q_0^\prime + \int_{\tau_0}^\tau\frac{{\rm d}\tau^{\prime}}
                                     {3f_{{\rm d}B}(\tau^{\prime})} - \tau\\
    &=& q_0 + \int_{\tau_0}^\tau\left[\frac{1}
         {3f_{{\rm d}B}(\tau^{\prime})}-1\right]{\rm d}\tau^{\prime} \ ,
         \label{qqrad2}
\eea
with $q_0 = q_0^\prime + \tau_0$.
That last equality, Eq.~(\ref{qqrad2}), constitutes the Hopf
function for radiative equilibrium. Similarly we find the Hopf function for the
partially convective atmosphere to be
\be{qqconv}
    q_{\rm conv}(\tau)
     = q_0 + \int_{\tau_0}^\tau\left[\frac{f_{\rm rad}(\tau^{\prime})}
         {3f_{{\rm d}B}(\tau^{\prime})}-1\right]{\rm d}\tau^{\prime}\ ,
\ee
where we introduced the fraction,
$f_{\rm rad} = F_{\rm rad}/F_{\rm tot}$, of the total flux,
$F_{\rm tot}=\sigma T_{\rm eff}^4$, which is carried by radiation.

In practise, we assume there is no convective flux at the upper boundary of the
simulations (it is less than \hbox{$3\stimes 10^{-5}F_{\rm tot}$} for all our
simulations), and simply use the average temperature at $\tau_0$ for
the integration constant,
\be{qq0}
    q_0 = \frac{4}{3}\left(\frac{T(\tau_0)}{T_{\rm eff}}\right)^4 - \tau_0\ ,
\ee
which is therefore the same for Eqs.~(\ref{qqrad2}) and (\ref{qqconv}).

The use for {\Ttau}s in the grey case is widely appreciated, but we have now
included convective effects and put the whole formulation on a firmer footing.
The final step is to ensure that they can be used in the general non-grey case
and therefore provide a valid and useful description of real stellar
atmospheres.

\subsection{$T(\tau)$-Relations from Non-Grey Atmospheres}
\label{AvRadTrans}

If the opacity {\em does} depend on wavelength,
integrating Eq.~(\ref{rad1}) over wavelength gives
\be{Ross1}
    \int_0^\infty{1\over \varrho\kappa_\lambda}\dd{K_\lambda}{z}{\rm d}\lambda
         = H = \frac{F_{\rm rad}}{4\pi}\ ,
\ee
where we substituted ${\rm d}\tau_\lambda=\varrho\kappa_\lambda{\rm d}z$.
$F_{\rm rad}$ and $H$ are just the results of direct integration over
wavelength.

The definition of the Rosseland mean opacity is motivated by the desire for
an average opacity that can be taken outside the integral of Eq.~(\ref{Ross1}).
This goal can be further simplified by noting that
$3K_\lambda\rightarrow J_\lambda\rightarrow S_\lambda$ for
$\tau_\lambda\rightarrow\infty$ and that in LTE $S_{\lambda}=B_{\lambda}$.
This results in
the usual definition of the {\em Rosseland opacity}
\be{RossOp}
    \frac{1}{\kappa_{\rm Ross}} =
    \frac{\int_0^\infty{\kappa_\lambda^{-1}}({\rm d}B_\lambda/{\rm d}T){\rm d}\lambda}
         {\int_0^\infty({\rm d}B_\lambda/{\rm d}T){\rm d}\lambda}\ .
\ee
(The differentiation with respect to $z$ in Eq.~(\ref{Ross1})
and $T$ in Eq.~(\ref{RossOp}) can be freely interchanged, if
they are monotonic functions of each other.)

In order to cast Eq.~(\ref{Ross1}) in a form similar to the grey version of
Eq.~(\ref{rad1}), ${\rm d}K/{\rm d}\tau = H$, we now define
$\widetilde K$ by
\be{Kross}
    \dd{\widetilde{K}}{\tau_{\rm Ross}} \equiv
        \kappa_{\rm Ross}\int_0^\infty{1\over\kappa_\lambda}
                        \dd{K_\lambda}{\tau_{\rm Ross}}{\rm d}\lambda
        = H = \frac{F_{\rm rad}}{4\pi}\ .
\ee

In the non-grey case the tight link between the moments of the intensity and
the moments of the transfer equation, is broken, as $K\ne\widetilde{K}$ in general.
Before we can use Eq.~(\ref{Kross}) as a basis for computing
$f_{\d B} = {\rm d}\widetilde{K}/{\rm d}B$,
we need to
ensure convergence in the radiative case. Dividing Eq.~(\ref{Kross}) by
$\d B/\d\tau_{\rm Ross}$ to obtain $f_{\d B}$, expanding $\kappa_{\rm Ross}$ of
Eq.~(\ref{Kross}), and finally exchanging the remaining differentiations by
$\tau_{\rm Ross}$ for ones by $T$ (which can be done since they are monotonic
functions of each other), we get
\be{fdBcnvrg}
    f_{\d B} = 
    \frac{\int_0^\infty{\kappa_\lambda^{-1}}(\d K_\lambda/\d T)\d\lambda}
         {\int_0^\infty{\kappa_\lambda^{-1}}(\d B_\lambda/\d T)\d\lambda}\ .
\ee
Since we know the ratio to converge for each wavelength, according to
Eq.~(\ref{fdB}) (which is for the grey case, but also valid monochromatically),
the integrals must similarly converge. For each wavelength the ratio will
converge as in the grey case, but in evaluating the total, spectral lines will
spread the transition over a larger range of heights.
With all the steps in Eq.~(\ref{fdB}) now validated for the non-grey case
(using $\widetilde{K}$ instead of $K$),
$f_{\d B}$ will in practise be evaluated as
\be{fdBRoss}
  f_{{\rm d}B}(\tau_{\rm Ross}) 
  = \frac{H(\tau_{\rm Ross})}{{\rm d}B/{\rm d}\tau_{\rm Ross}}
  = \frac{1}{16\sigma T^3}\frac{F_{\rm rad}}{{\rm d}T/{\rm d}\tau_{\rm Ross}}\ ,
\ee
and temperatures can be found from Eqs.~(\ref{qqconv}) and (\ref{qqrad}) using
$\tau=\tau_{\rm Ross}$. This is the procedure for computing radiative {\Ttau}s
from 1D atmospheres.

From this analysis we conclude that {\Ttau}s
are perfectly suited for describing real stellar atmospheres.
Since the formulation does not rely on $S=B$, except in the optically deep layers,
this formulation will also be valid for non-LTE atmospheres, whether they
just include continuum scattering or a thorough implementation of non-LTE
effects. The choice of $\kappa_{\rm Ross}$ as the standard opacity is merely to
ensure proper convergence with depth and as a convenience to stellar modellers.
This does not limit the scope of the
formulation in the photosphere and above.
It is often asserted that the use of {\Ttau}s with a Rosseland opacity
amounts to using the grey approximation. We hope that we have demonstrated
this to be far from the case and that {\Ttau}s can describe arbitrarily
realistic and complex atmospheres.
The use of a Rosseland opacity to reconstitute the atmosphere in structure
models says nothing about the level of complexity that went into the original
atmosphere calculation from which the {\Ttau} was derived. Reconstituting the
{\Ttau} with the Rosseland opacity results in the structure of the original
atmosphere, but without the need to perform the computationally expensive
radiative transfer calculation.

\subsection{$T(\tau)$-Relations in 3D}
\label{av-proc}

In 1D there is no ambiguity about the meaning of a certain quantity, {\eg},
$\tau$ and $T$, and they depend in simple ways on the height in the atmosphere,
$z$. In 3D, on the other hand, there is a whole range of temperatures and
optical depths at a particular height, and different averaging methods will
give rather different results. This is illustrated in Fig.~\ref{mT-tau} where
we compare three different averaging methods 
for our solar simulation (No.\ 30 in table \ref{simlist}).
We plot temperatures averaged over the undulated surfaces of equal optical depth
($\tau$-average) in case {\bf a)}. In case {\bf b)} we plot a straight horizontal average
of the temperature as function of a straight horizontal average of the optical
depth, $\langle\dots\rangle_z$. For the temporal averaging of these, we map the
horizontal averages to a fixed column mass scale. We refer to these as pseudo
(since we use the average, not the local column mass) Lagrangian averages,
denoted by $\langle\dots\rangle_{\rm L}$. This filters out the main effect of
the p modes that are excited in the simulations. Finally,
in case {\bf c)} we show the horizontally averaged temperature as function of an
optical depth integrated from the opacity of the horizontally averaged density
and temperature.
\mfigur{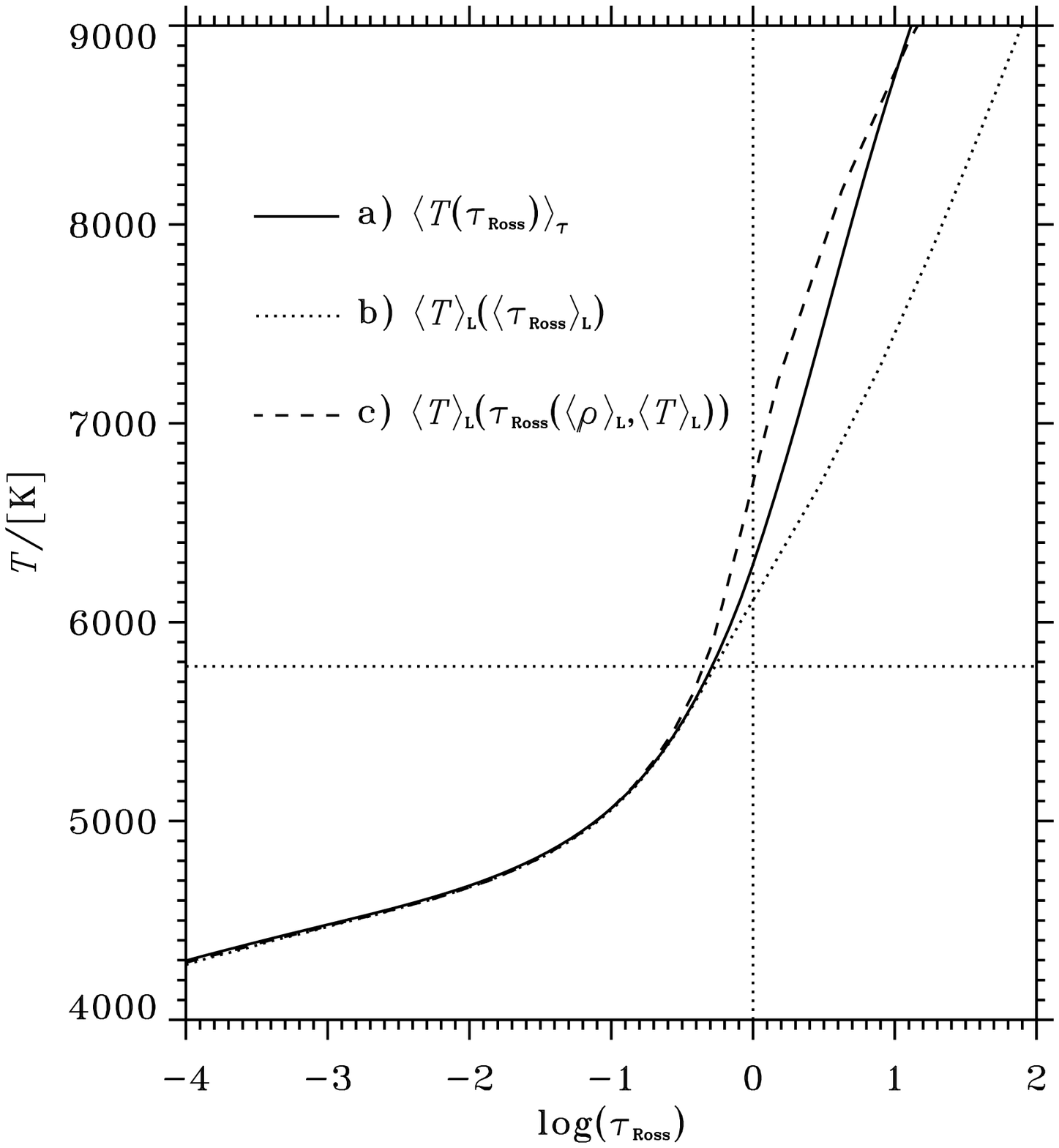}{9.0cm}
    {The effect of various averaging techniques
     applied to a solar simulation (see text for details).
     Notice how cases a) and b) follow closely in the atmosphere down to
     $\langle T\rangle=T_{\rm eff}$, at $\log\tau\simeq -0.3$. The horizontal
     dotted line indicates $T_{\rm eff}$ and the vertical, unity optical depth.
     The solar {\Ttau}s in this plot include the first-order convective effect.}

Not unexpectedly, these three different methods diverge in the region where the temperature
fluctuations are the largest, from the photosphere down to 1\,Mm below (at
$\tau_{\rm Ross}\sim 10^5$). There the convective fluctuations, in mainly the
temperature, are so large that the opacity and the EOS ({\eg}, gas pressure,
$p_{\rm g}$) are non-linear on the scale of the fluctuations. This has the
consequence that
$\langle p_{\rm g}\rangle \neq p_{\rm g}(\langle\varrho\rangle, \langle T\rangle)$.
In other words, the average gas
pressure is, in general, not related to the average density and temperature
through the EOS. The relative difference amounts to about 5\% in the solar
photosphere. This effect is much larger for the opacity where the relative
difference reaches more than 90\% in the solar photosphere. Well above the photosphere, the averaging
methods converge, as the temperature fluctuations decrease, and more
importantly, as the opacity varies less steeply.

The large differences between the methods around the photosphere mean that
great care is needed in choosing the appropriate form of averaging.
Intuitively quantities having to do with radiative transfer would be best
represented as $\tau$-averages, case a) above.
This choice was also advocated by \citet{ludwig:alfa-cal}, who used
$\tau$-averaged temperatures of their 2D convection simulations as {\Ttau}s
for corresponding 1D envelope models. These then formed the basis for a
calibration of the mixing-length, similar to what we present in Paper II.

Averaging on the $\tau$-scale, however, presents problems for the concept of
slanted rays through the atmosphere, and hence for the angular moments of the
intensity. On the $\tau$-scale these rays will no longer be straight lines
through the (inclined) simulation box,
and the formulation becomes impractical, rendering the straight horizontal
average the obvious choice. We therefore
need to rephrase the transfer equation, Eq.~(\ref{transfer}), and its first
angular moment, Eq.~(\ref{Kross}), in terms of $z$ instead of $\tau$.
This is option {\bf b)}, mentioned above, where $\langle\dots\rangle_z$ are
functions of time. Temporal, pseudo-Lagrangian averaging,
$\langle\dots\rangle_{\rm L}$, is performed on the final products of our
derivation, the radiative Hopf functions.
Recasting in terms of $\tau$ results in
\be{dKdz}
    \dd{K}{z} = \int_0^\infty \dd{K_\lambda}{z}{\rm d}\lambda
              = \varrho\int_0^\infty\kappa_\lambda H_\lambda{\rm d}\lambda
              \equiv \widetilde{H}\ ,
\ee
which is used to form our invariant
\be{fdBz}
    f^z_{{\rm d}B} = \frac{\langle{\rm d}K\rangle_z}{\langle{\rm d}B\rangle_z}
                 = \frac{\langle{\rm d}K/{\rm d}z\rangle_z}
                         {\langle{\rm d}B/{\rm d}z\rangle_z}
                 = \frac{\langle\widetilde{H}\rangle_z}
                        {{\rm d}\langle B\rangle_z/{\rm d}z}\ ,
\ee
with $\langle\dots\rangle_z$ denoting horizontal averages.
This 3D version of $f^z_{{\rm d}B}$ converges as shown for the 1D case
in Sect.~\ref{AvRadTrans}, although this convergence also depends on the
horizontal fluctuations becoming insignificant with depth, so that the 1D
version, Eq.~(\ref{fdBRoss}), applies again. We have confirmed, numerically,
that $f^z_{{\rm d}B} \rightarrow 1/3$ well before the bottom of the simulation
domains. Due to the increasing numerical instability with depth (since 
$f^z_{{\rm d}B}$ is a ratio between exponentially increasing quantities), we
actually enforce an exponential convergence to 1/3, from just before
the first point that
goes above 1/3 (due to numerical noise in the hydrodynamics and the radiative
transfer). The change to $f^z_{{\rm d}B}$ due to this forced convergence,
is small and not systematic, but it greatly improves the convergence and smoothness of
the resulting $q_{\rm rad}$ below the photosphere.

Similar to what we did for Eq.~(\ref{Bint}), we integrate
$\widetilde H/f^z_{{\rm d}B}$ in order to obtain $B_{\rm rad}(z)$ and hence the
radiative Hopf function, as in Eq.~(\ref{qqconv}),
\be{qrad3D}
    q_H^{\rm rad}(z) = q_0 + \int_{z_0}^z\left[\
    \frac{4\pi}{3}\frac{\langle\widetilde H\rangle/F_{\rm rad}}
       {f^z_{{\rm d}B}} - \langle\varrho\kappa_H\rangle_z\right]{\rm d}z^\prime\ ,
\ee
where the $\tau_H$-scale, defined by $\langle\varrho\kappa_H\rangle_z$
in terms of the $H$-averaged opacity $\kappa_H$,
remains to be determined.
In order for the radiative Hopf function
to be convergent for $\tau_H\gg 1$, the square bracket must converge to 0.
With the convergent $f^z_{{\rm d}B} \rightarrow\frac{1}{3}$, we must therefore
have
\be{qrad3D2}
q_H^{\rm rad}(z) = q_0 + \int_{z_0}^z \langle \varrho\kappa_H\rangle\left[
              \frac{1}{3f^z_{{\rm d}B}} - 1\right]{\rm d}z^\prime\ ,
\ee
for $\tau_H\gg 1$. This defines $\kappa_H$ such that it satisfies
\be{abH}
    \langle\varrho\kappa_H\rangle_z(z) \equiv
        4\pi\frac{\langle\widetilde H\rangle_z(z)}{F_{\rm rad}(z)}
      = \frac{\langle\widetilde H\rangle_z(z)}{\langle H\rangle_z(z)}
      = \dd{\tau_H}{z}\ .
\ee
In practise, $q^{\rm rad}_H$ is fully converged at $\tau_H \simeq 10$.

We further demand consistency, so that the full (partially convective) Hopf function,
$q^{\rm conv}_H(z)$, defined by
\be{qtot3D}
\frac{4}{3}\left(\frac{T(z)}{T_{\rm eff}}\right)^4 = 
    q^{\rm conv}_H(z) + \tau_H = q_0 + \int_{z_0}^z\frac{4\pi}{3}   
        \frac{\langle{\widetilde H}\rangle/F_{\rm tot}}
                                            {f^z_{{\rm d}B}}{\rm d}z^\prime\ ,
\ee
can reproduce the actual temperature structure of the simulation. With
$\langle{\widetilde H}\rangle$ cancelling from Eq.~(\ref{fdBz})
for $f^z_{{\rm d}B}$,
$B=\sigma T^4/\pi$ and $F_{\rm tot}=\sigma T_{\rm eff}^4$, we see that this
is indeed fulfilled, and that
$q_0 = \mfrac{4}{3}\langle T^4\rangle(z_0)/T_{\rm eff}^4 - \tau_H(z_0)$
as in Eq.~(\ref{qq0}).

The $\tau_H$-scale defined by Eq.~(\ref{abH}) is a bit awkward, since it
demands knowledge of the complete radiative transfer solution - something we
explicitly are trying to avoid with our approach. We therefore want to rephrase
$q$
in terms of the ``universal'' Rosseland opacity, which depends only on the
local thermodynamic state.

Since the first term in the integrand of $q(z)$ in Eq.\ (\ref{qrad3D})
is independent
of the $\tau$-scale and the second term is just the $\tau$-scale itself
this can be achieved by replacing $\langle \varrho \kappa_H \rangle$
by $\langle \varrho \kappa_{\rm Ross} \rangle$,
for a suitably defined $\kappa_{\rm Ross}$, to obtain
\bea{qRoss}
    q_{\rm Ross}^{\rm rad}(z) {\nsp}&=&{\nsp} q_0 + \int_{z_0}^z\left[\
        \frac{4\pi}{3}\frac{\langle\widetilde H\rangle/F_{\rm rad}}
           {f^z_{{\rm d}B}} - \langle\varrho\kappa_{\rm Ross}\rangle_z\right]
         {\rm d}z^\prime \\
     {\nsp}&=&{\nsp} q_0 + \int_{\tau_{\rm Ross,0}}^{\tau_{\rm Ross}}\left[\
        \frac{4\pi}{3}\frac{\langle\widetilde H\rangle/F_{\rm rad}}
           {f^z_{{\rm d}B}\langle\varrho\kappa_{\rm Ross}\rangle_z} - 1 \right]
         {\rm d}\tau_{\rm Ross}^\prime \; ,
        \label{qRoss2}
\eea
which defines the radiative Hopf function on the Rosseland scale.
However, unless $\kappa_{\rm Ross}\rightarrow\kappa_H$ for $\tau\gg 1$,
then $q_{\rm Ross}^{\rm rad}(z)$ will not converge to a constant either.

Unfortunately,
what would normally be interpreted as the average Rosseland opacity
\emph{does not} converge to
the flux averaged opacity, $\langle\varrho\kappa_H\rangle_z$,
as defined by Eq.~(\ref{abH}). On the other hand, $q(\tau_H)$ depends on
this definition in order to fulfil the constraints of converging $f^z_{{\rm
d}B}$ and $q_{\rm rad}(\tau_H)$ and the ability of $q_{\rm conv}(\tau_H)$ to reproduce
the temperature stratification of the simulations.

The solution to this problem is to cast the averaging method for
the Rosseland opacity into a form similar to Eq.~(\ref{abH}),
where the opacity and the normalizer
(see Eq.~[\ref{RossOp}]) are averaged separately
\be{mRossOp1}
    {\left\langle\dd{B}{T}\right\rangle_z}\left/
      {\left\langle\int_0^\infty(\kappa_\lambda^{-1}
                    \dd{B_\lambda}{T}){\rm d}\lambda\right\rangle_z}\right.\ ,
\ee
but by using ${\rm d}B/{\rm d}z$ instead, resulting in
\be{mRossOp2}
    \langle\varrho\kappa_{\rm Ross}\rangle_z \equiv
    {\left\langle\dd{B}{z}\right\rangle_z}\left/
   {\left\langle\frac{1}{\varrho\kappa_{\rm Ross}}\dd{B}{z}\right\rangle_z}\right.\ ,
\ee
to be used in Eq.\ (\ref{qRoss}).
The $\langle\varrho\kappa_H\rangle$ of Eq.~(\ref{abH}) converges properly to
this form of the horizontally averaged Rosseland opacity for $\tau\gg 1$.
This can be ascertained by relating the numerators of Eq.\ (\ref{abH}) and
(\ref{mRossOp2}) through Eq.\ (\ref{dKdz}), and separately their denominators
through Eq.\ (\ref{Ross1}), and realising that both sets converge to each other
as $3K_\lambda\rightarrow B_\lambda$.

In and above the photosphere the two opacities diverge, as the diffusion approximation no
longer holds,
%
%
$\kappa_H$ being the larger of the two (since the flux spectrum is largely
determined at the photosphere and the temperature there, whereas
${\rm d}B_\lambda/{\rm d}\tau_\lambda$ is entirely set by the local temperature.
This red-shifts the Rosseland weighting function towards the lower opacity of
the H$^-$ bump, away from the crowded lines and metallic absorption edges in
the UV). This means
$q_{\rm Ross} > q_H$, and in the optically deep layers they are merely offset
by a constant.
It is also worth noting
that the 1D form of Eq.~(\ref{mRossOp2}) is identical to the conventional form,
leaving our formulation consistent with previous 1D work.

\subsection{Implementation of the $T(\tau)$-Relation in Stellar Structure Models}
\label{Ttau-impl}

We proved above that {\Ttau}s can describe real stellar atmospheres and
therefore have the potential to provide the outer boundary conditions of
stellar structure models. In this section we will show in detail how this
is carried out, in particular how convective effects are reintroduced in a
consistent manner. The reason for this elimination and subsequent reintroduction
of convection to the {\Ttau} is the resulting independence of convection
treatment between the atmosphere models (providing the {\Ttau}) and the
stellar structure models (employing the {\Ttau}). Thus, the models can be
internally self-consistent, despite differing greatly in their convection treatment
-- in our case 3D convection simulations and 1D with the mixing-length 
\citep[MLT]{boehm:mlt} formulation of convection.

From here on we will use the abbreviation $q=q_{\rm Ross}^{\rm rad}$,
and a Rosseland mean is implied for both opacity and optical depth, unless
explicitly stated otherwise. All quantities in this section pertain to the
structure model implementing the {\Ttau}, except for $q(\tau)$ which, of
course, is computed from the atmosphere model.

In deriving the {\Ttau} in Sect.~\ref{grey} and \ref{AvRadTrans}
we made the transformation from
actual to radiative {\Ttau} by changing $T$, as shown in Eq.~(\ref{qqrad}).
In the envelope calculations we need to re-introduce convection in the {\Ttau},
this time described by, {\eg}, MLT.
Defining
\be{taueff}
    {\rm d}\hat\tau = f_{\rm rad}{\rm d}\tau\ ,
\ee
we see that the full temperature profile, from Eq.~(\ref{qtot3D}), can be
rewritten to give
%
\bea{qtauhat}
    \frac{4}{3}\left(\frac{T(\tau)}{T_{\rm eff}}\right)^4
{\!\!\nsp}&=&{\nsp} q_0 + \int_{z_0}^z\left[\frac{4\pi}{3}
        \frac{\langle\widetilde H\rangle/F_{\rm tot}}
             {f_{{\rm d}B}^z}-\langle\varrho\kappa\rangle\right]
                                            {\rm d}z^\prime + \tau\\
          &=&{\nsp} q_0 + \int_{\tau_0}^\tau\left[\frac{4\pi}{3}
        \frac{\langle\widetilde H\rangle/F_{\rm tot}}
             {\langle\varrho\kappa\rangle f_{{\rm d}B}^z}-1\right]
                    \label{qtauhat1}        {\rm d}\tau^\prime + \tau\\
          &=&{\nsp} q_0 + \int_{\hat\tau_0}^{\hat\tau}\left[\frac{4\pi}{3}
        \frac{\langle\widetilde H\rangle/F_{\rm rad}}
             {\langle\varrho\kappa\rangle f_{{\rm d}B}^z}-1\right]
                                            {\rm d}\hat\tau^\prime + \hat\tau\\
          &\simeq&{\nsp} q(\hat\tau) + \hat\tau \label{qtauhat2}\ .
\eea
The last line results in the very simple transformation
\be{Hopfhat}
    q_{\rm rad}(\hat\tau) + \hat\tau = q_{\rm conv}(\tau) + \tau\ ,
\ee
entirely accomplished through a modification of the optical depth.
Equation (\ref{qtauhat2}) is not exact, since the first term of the integrand
is evaluated at $\tau$ and not $\hat\tau$, as implied by Eq.~(\ref{qtauhat2}).
For all the cases we have dealt with
the differences are small, corresponding to a less than 0.2\,K increase of
the temperature in and above the photosphere, spanning about two orders of magnitude in optical
depth.

The right-hand side is the complete {\Ttau} of the structure model, including
the transition to (1D)
convection, and $q_{\rm rad}(\tau) + \tau$ is the radiative {\Ttau} from the detailed
atmosphere model,
converging properly for $\tau \rightarrow \infty$.
The change of $\tau \rightarrow \hat\tau$ on the left-hand side
is solely responsible for recovering
the partially convective {\Ttau} from a
radiative Hopf function.
Due to the simple behaviour of the first-order effect of convection on the
temperature stratification (a consequence of Eq.~[\ref{rad1}]), the radiative
equilibrium stratification and the actual stratification can be described by
the same Hopf function, merely by a simple change to the argument.

The radiative temperature gradient is, as usual, defined
as the gradient that would be caused by radiative transport of energy alone,
{\ie}, assuming that $T$ is given by $q(\tau)+\tau$. Differentiating with
respect to $r$, using ${\rm d}\tau=-\varrho\kappa{\rm d}r$, and dividing by the
equation of hydrostatic equilibrium, ${\rm d}p/{\rm d}r =-g\varrho$, we find
\be{nabrad}
    \nabla_{\rm rad} \equiv \left({{\rm d} \ln T \over {\rm d} \ln p} \right)_{\rm rad}
                         = {3\over 16\sigma}{\kappa F_{\rm tot}p\over gT^4}
                           \left[q^\prime(\tau) + 1 \right]\ ,
\ee
($\ln$ being the natural logarithm)
where $p$ is the total hydrostatic pressure, possibly including a turbulent
contribution from the vertical component of the convective velocity field,
$p_{\rm turb} = \langle\varrho u_z^2\rangle$ (see Paper\,II for a discussion
of $p_{\rm turb}$).

The actual gradient, $\nabla$, can similarly be found by using
the left hand side of Eq.~(\ref{Hopfhat}) so that,
\be{nab}
    \nabla \equiv  \dd{\ln T}{\ln p}
                         = {3\over 16\sigma}{\kappa F_{\rm tot}p\over gT^4}
                       \dd{}{\tau}\left[q(\hat\tau) + \hat\tau\right]\ .
\ee
Cancelling the two $\tau$-terms in Eq.~(\ref{qtauhat1}) we find the derivative
in Eq.~(\ref{nab}) to be the first term of the integrand of
Eq.~(\ref{qtauhat1})
\be{dqhatdtau}
    \dd{}{\tau}\left[q(\hat\tau) + \hat\tau\right]
        = \frac{4\pi}{3}\frac{\langle\widetilde H\rangle/F_{\rm tot}}
                             {\langle\varrho\kappa\rangle f_{{\rm d}B}^z}\ ,
\ee
(all evaluated at $\tau$) which we also recognise, using
Eq. (\ref{qRoss2}), as
$\left[q^\prime(\tau) + 1 \right]f_{\rm rad}$.
This results in
\be{fkappa2}
    \frac{\nabla}{\nabla_{\rm rad}}
        = \dd{\hat\tau}{\tau} = f_{\rm rad} \ .
\ee
Had we taken the derivative of the approximate expression, Eq.~(\ref{qtauhat2}),
the result would have been more complicated. With the small effect of the
approximation, we have seen no problems arising from the slight inconsistency.

With these relations the {\Ttau} can be used throughout the stellar
envelope model, without the (common) artificial transition between atmosphere
and interior.
A somewhat similar, but less rigorous, approach was used by \citet{henyey:stel-evol3}
to ensure a smooth transition between the atmosphere and interior
of stellar models.

\subsection{Summary of Procedures}
\label{recipe}

As derived above, there are two separate steps involved: First the calculation
of the {\Ttau}, in the form of a generalised Hopf function $q(\tau)$, from
an atmosphere model which explicitly solves the radiative transfer equation.
Second, the implementation of {\Ttau}s in 1D stellar structure models.

The purely radiative Hopf function, $q(\tau)$, is computed from 1D stellar
atmospheres
from Eq.\ (\ref{qqrad2}) and the invariant $f_{{\rm d}B}$ of
Eq.\ (\ref{fdBRoss}), both based on the Rosseland opacity.

For 3D atmospheres the procedure is slightly more complicated, although it
reduces properly to the 1D case above. In 3D the invariant, $f_{{\rm d}B}$,
is computed from
Eq.\ (\ref{fdBz}) and applied in Eq.\ (\ref{qRoss}) for the radiative Hopf
function. The horizontally averaged Rosseland opacity,
$\langle\varrho\kappa_{\rm Ross}\rangle_z$, needs to be slightly
redefined, so that the wavelength integrals of the opacity and of the
normalization factor are averaged separately, as specified by
Eq.\ (\ref{mRossOp2}).

The implementation of the {\Ttau} is accomplished through modifications of
the radiative, $\nabla_{\rm rad}$, and actual temperature gradients, $\nabla$,
Eqs.\ (\ref{nabrad}) and
(\ref{nab}), respectively. The actual gradient involves the radiative Hopf
function evaluated at a modified $\tau$-scale, $q(\hat\tau)$, constructed to
account for the first order convective effect on the stratification, due to
departure from radiative equilibrium. This modification is defined by
Eq.\ (\ref{fkappa2}).

We have provided electronic tables of $q(\tau)$ for our
grid of simulations, and of the [Fe/H]=0.0 Rosseland opacity,
as well as code for reading and interpolating in these
tables. 

\section{The 3D Convection Simulations}
\label{sims}

The fully compressible, transmitting-boundary,
radiation-coupled hydrodynamic (RHD) simulations that we rely on here are
described by \citet{trampedach:3Datmgrid}.
A recent and thorough review of applications of such 3D stellar convection simulations can be found in
\citet{nordlund:LivingRev}.

The EOS and the
opacities were originally supplied by the so-called Uppsala package
by \citet{b.gus} which forms the atomic-physics basis for the original
MARCS atmosphere models \citep[greatly improved
and updated in the present MARCS models by \citealt{MARCS-2}]{gus:modgrid}.
Since the matching to 1D envelopes (employed in the mixing-length calibration
of Paper II) requires a high degree of consistency
between the simulations and the envelopes, we found it necessary to
bring the micro-physics of the 3D simulations up to the same level as that of
the 1D envelopes.
The ever stronger constraints from improving observations also demanded
an upgrade to more realistic atomic physics.

We therefore implemented the realistic
Mihalas-Hummer-D{\"a}ppen (MHD) EOS \citep{mhd1,mhd3}.
This EOS treats hundreds of bound levels explicitly for every ionization level
of every included element. The hydrogen molecules H$_2$ and H$_2^+$
are
included as the only molecules, and non-ideal interactions leading to pressure
ionization are treated in detail, as are degenerate electrons.
Furthermore, thermodynamic consistency (that the Maxwell relations are obeyed)
is ensured by the use of the free energy minimisation procedure
\citep{wd:fre-e-mini}, which was also a motivation for the EOS update.
The EOS tables were custom calculated for a 15 element mixture (as opposed to the normal 6), to correspond
to what is included in the Uppsala package. 
\aatab{simlist}{Fundamental parameters for the 37 simulations.}{rccccl}
{ sim & MK\,class & \oh{$T_{\rm eff}$/[K]}& \oh{$\log g$}& \oh{$M/M_\odot$} & star \\}{
  1 &     K3 & $4\,681\pm 19$ & 2.200 & 3.694 &                   \\
  2 &     K2 & $4\,962\pm 21$ & 2.200 & 4.805 &                   \\
  3 &     K5 & $4\,301\pm 17$ & 2.420 & 0.400 &                   \\
  4 &     K6 & $4\,250\pm 11$ & 3.000 & 0.189 &                   \\
  5 &     K3 & $4\,665\pm 16$ & 3.000 & 0.852 &                   \\
  6 &     K1 & $4\,994\pm 15$ & 2.930 & 2.440 & $\xi$\,Hya        \\
  7 &     G8 & $5\,552\pm 17$ & 3.000 & 2.756 &                   \\
  8 &     K3 & $4\,718\pm 15$ & 3.500 & 0.721 &                   \\
  9 &     K0 & $5\,187\pm 17$ & 3.500 & 1.786 &                   \\
 10 &     K0 & $5\,288\pm 20$ & 3.421 & 1.923 & $\nu$\,Ind        \\
 11 &     F9 & $6\,105\pm 25$ & 3.500 & 1.875 &                   \\
 12 &     K6 & $4\,205\pm ~8$ & 4.000 & 0.601 &                   \\
 13 &     K4 & $4\,494\pm ~9$ & 4.000 & 0.684 &                   \\
 14 &     K3 & $4\,674\pm ~8$ & 4.000 & 0.738 &                   \\
 15 &     K2 & $4\,986\pm 13$ & 4.000 & 0.836 &                   \\
 16 &     G6 & $5\,674\pm 16$ & 3.943 & 1.130 & $\beta$\,Hyi      \\
 17 &     F9 & $6\,137\pm 14$ & 4.040 & 1.222 &                   \\
 18 &     F4 & $6\,582\pm 26$ & 3.966 & 1.567 & Procyon           \\
 19 &     F4 & $6\,617\pm 33$ & 4.000 & 1.542 &                   \\
 20 &     K4 & $4\,604\pm ~8$ & 4.300 & 0.568 &                   \\
 21 &     K1 & $4\,996\pm 17$ & 4.300 & 0.694 &                   \\
 22 &     K1 & $5\,069\pm 11$ & 4.300 & 0.719 &                   \\
 23 &     K0 & $5\,323\pm 16$ & 4.300 & 0.810 &                   \\
 24 &     G1 & $5\,926\pm 18$ & 4.295 & 1.056 & $\alpha$\,Cen\,A  \\
 25 &     F5 & $6\,418\pm 26$ & 4.300 & 1.261 &                   \\
 26 &     F2 & $6\,901\pm 29$ & 4.292 & 1.433 &                   \\
 27 &     K4 & $4\,500\pm ~4$ & 4.500 & 0.565 &                   \\
 28 &     K3 & $4\,813\pm ~8$ & 4.500 & 0.664 &                   \\
 29 &     K0 & $5\,232\pm 12$ & 4.500 & 0.812 &                   \\
 30 &     G5 & $5\,774\pm 17$ & 4.438 & 1.002 & The Sun           \\
 31 &     F7 & $6\,287\pm 15$ & 4.500 & 1.246 &                   \\
 32 &     F4 & $6\,569\pm 17$ & 4.450 & 1.329 &                   \\
 33 &     K1 & $5\,021\pm 11$ & 4.550 & 0.772 &                   \\
 34 &     G9 & $5\,485\pm 14$ & 4.557 & 0.949 & $\alpha$\,Cen\,B  \\
 35 &     G1 & $5\,905\pm 15$ & 4.550 & 1.114 &                   \\
 36 &     K6 & $4\,185\pm ~3$ & 4.740 & 0.649 &                   \\
 37 &     K4 & $4\,531\pm 10$ & 4.740 & 0.742 &                   \\
}T

Improvements to the continuum opacities are described in Sect.\ \ref{sect:opac}.
Line opacity is supplied by the opacity distribution functions (ODFs) of
\citet{kur:line-data,kur:missolar}, and
line blanketing is included in the radiative transfer by means of
opacity binning \citep{aake:numsim1}. The spectrum is divided into four bins according to the
strength of opacity at each wavelength, and the source function is summed-up
for each bin, as described in detail by 
\citet{trampedach:thesis} and \citet{trampedach:3Datmgrid}. Radiative transfer in the simulations is then
calculated for just these four bins, reducing the problem by more than three
orders
of magnitude, and making it tractable in 3D. This binning scheme does not
benefit from increasing the number of bins beyond four. From comparisons with
the full monochromatic radiative transfer we estimate that the temperature
error due to this binning is less than 40\,K for the solar case, in the range
of $-4 < \log\tau < 0$. For each bin the transfer
equation, Eq.~(\ref{transfer}), is solved for the vertical ray and four
slanted, $\mu=\sfrac{1}{3}$ rays, equally spaced in azimuthal angle (giving a
total of five rays). The azimuthal dimension is of course degenerate in 1D,
but adds another dimension to the 3D problem.
The bin assignment is based on a full monochromatic
radiative transfer calculation for the temporally and horizontally averaged
simulation, and therefore varies between the simulations.
Detailed radiative transfer is
computed for all horizontal layers that have a minimum Rosseland optical depth
$\min(\tau) < 300$, and the diffusion approximation is used below.

Our radiative transfer is performed in strict LTE where $S=B$. We minimize the
effects of that approximation by excluding scattering cross sections from the
free-streaming, intensity-weighted opacity above the photosphere, and including
it as absorption in the Rosseland mean below. \citet{hayek:Parallel3Dscatter}
found from a consistent treatment of scattering that this is a good
approximation, at least for solar-metallicity atmospheres.

We have performed simulations for 37 sets of atmospheric parameters,
all with solar abundances (discussed below), as listed in Table~\ref{simlist}. This
table is ordered in ascending gravity and, for similar gravities, ascending
in effective temperature---the same order as in Table 2 of \citet{trampedach:3Datmgrid}. The spectral class is rather approximate, as is
the mass which is based on estimates of intersections with evolutionary tracks.
Neither of these two quantities are used in this work, but are merely included
for the reader's convenience. Some of the simulations correspond to actual
stars, as indicated in the right-most column, but with the caveat that all
the simulations have solar composition.

This set of simulations is the beginning of a grid of stellar
convection simulations, aimed at augmenting the existing grids of 1D stellar
atmosphere models. Other aspects and applications of the grid will be presented
in forthcoming papers.

For the solar composition we have chosen a helium mass fraction, $Y_\odot=0.245$, according to
helioseismic determinations \citep{basu-antia:Y}, and a metal to hydrogen ratio (by mass) of $Z_\odot/X_\odot=0.0245$ in agreement
with \citet{GN92}.
This ratio results in the hydrogen mass fraction, $X=0.736945$ and the
helium-hydrogen number ratio, $N({\rm He})/N({\rm H}) = 0.08370$, instead of the historically
assumed value of $\simeq 0.1$. Our metal mixture is constrained by that of the
available ODF tables \citep{kur:line-data} that were constructed for \citet{AG89} abundances.
Additional tables for different He and Fe abundances meant that we could
interpolate between tables to the He abundance above, $A({\rm He/H}) = 10.92$,
and $A({\rm Fe/H}) = 7.50$ from \citet{GN92}, where $A$ denotes logarithmic
number abundances, normalised to $A({\rm H}) = 12$.

The above solar composition has been challenged over the last decade
by abundance analyses
performed on 3D convection simulations using the same code and atomic physics as
used for the simulations in the present paper. 
The result is a general lowering of the metal abundances
as detailed by \citet[][and references therein]{AGS05}. Such a lower metal
abundance ruins the otherwise good agreement with helioseismic inversions,
as pointed out by \citet{basu:seism-AGS05} and \citet{bahcall:seism-AGS05}.
This controversy highlights the fact that our own star, the Sun, is less
well-known than often assumed and stresses the need for improved modelling
efforts as pursued with, {\eg}, the present work, and improved atomic physics
calculations \citep{rogers:newOPAL,OP05}.
The re-evaluation of the solar abundances by \citet{AGSS2009}, based on a new
solar simulation, has increased
the over-all metallicity by about 10\% and reduces the disagreement with
helioseismology to about 2/3 of the former \citep{serenelli:SunSeismAGSS09}.
The new simulation has improved radiative transfer and opacities
(Trampedach et {al.} in preparation) and agrees very well with observations of both
limb darkening, flux distribution and strong hydrogen lines
\citep{tiago:suneHlinesLimbdark}.

Due to the lingering controversy, and since the change in abundances will
have a rather small effect on stellar atmospheres, we did not feel compelled
to recompute the, at that time, mostly finished grid of simulations to adopt
the new abundances.

Each of the simulations was performed on a $150\stimes 150\stimes 82$-point
grid (equidistant horizontally and non-uniform vertically, optimized to resolve
the large photospheric gradients). This resolution is adequate for our purposes
and is guided by the study of resolution effects by \citet{asplund:num-res}.
The simulation domain covers about 10 major granules horizontally and 13 pressure scale heights
vertically, with about 20\% (in height) being above $\langle T\rangle=T_{\rm eff}$. We
ensured that Rosseland optical depths as low as $\log\tau=-4.5$ were completely
contained in each of the simulations. The area and depth of the solar simulation
is $6\stimes 6\stimes 3.6$\,Mm and the size of the other simulations is mainly
scaled by $g^{-1}$.
After
relaxation to a quasi-stationary state, we
calculated mean models as described
in Sect.~\ref{av-proc}. The relaxation involves a damping of radial p modes
which efficiently extracts surplus energy from the simulations. This damping
is turned off again for the production runs, 
on which our work is based.

\subsection{Opacities}
\label{sect:opac}

We have revised most of the continuum opacity sources and added a few more sources
as follows: H$^-$ bound-free (bf) \citep{broad-rein:H-,wish:H-},
free-free (ff) \citep{bell-berr:H-ff},
H$_2^+$ bf+ff \citep{stancil:H2+}, H$_2^-$ ff \citep{bell:H2}, and
OH/CH photo-dissociation \citep{kur:OH+CH}. The most important levels of
\ion{He}{i}, \ion{C}{i}, \ion{N}{i}, \ion{O}{i}, \ion{Na}{i}, \ion{Mg}{i},
\ion{Mg}{ii}, \ion{Al}{i}, \ion{Si}{i}, \ion{Ca}{i}, \ion{Ca}{ii} and
\ion{Fe}{i} were included as simple analytical fits as
listed by \citet{mathisen1}. The usual Thomson scattering by free
electrons is included, together with Rayleigh scattering by \ion{H}{i}
\citep{gavrila:Hray}, \ion{He}{i} \citep{Lang:Heray} and H$_2$ \citep{Vict-Dalg:H2ray}.
These changes were described in more detail by
\citet{trampedach:thesis}.
\mfigur{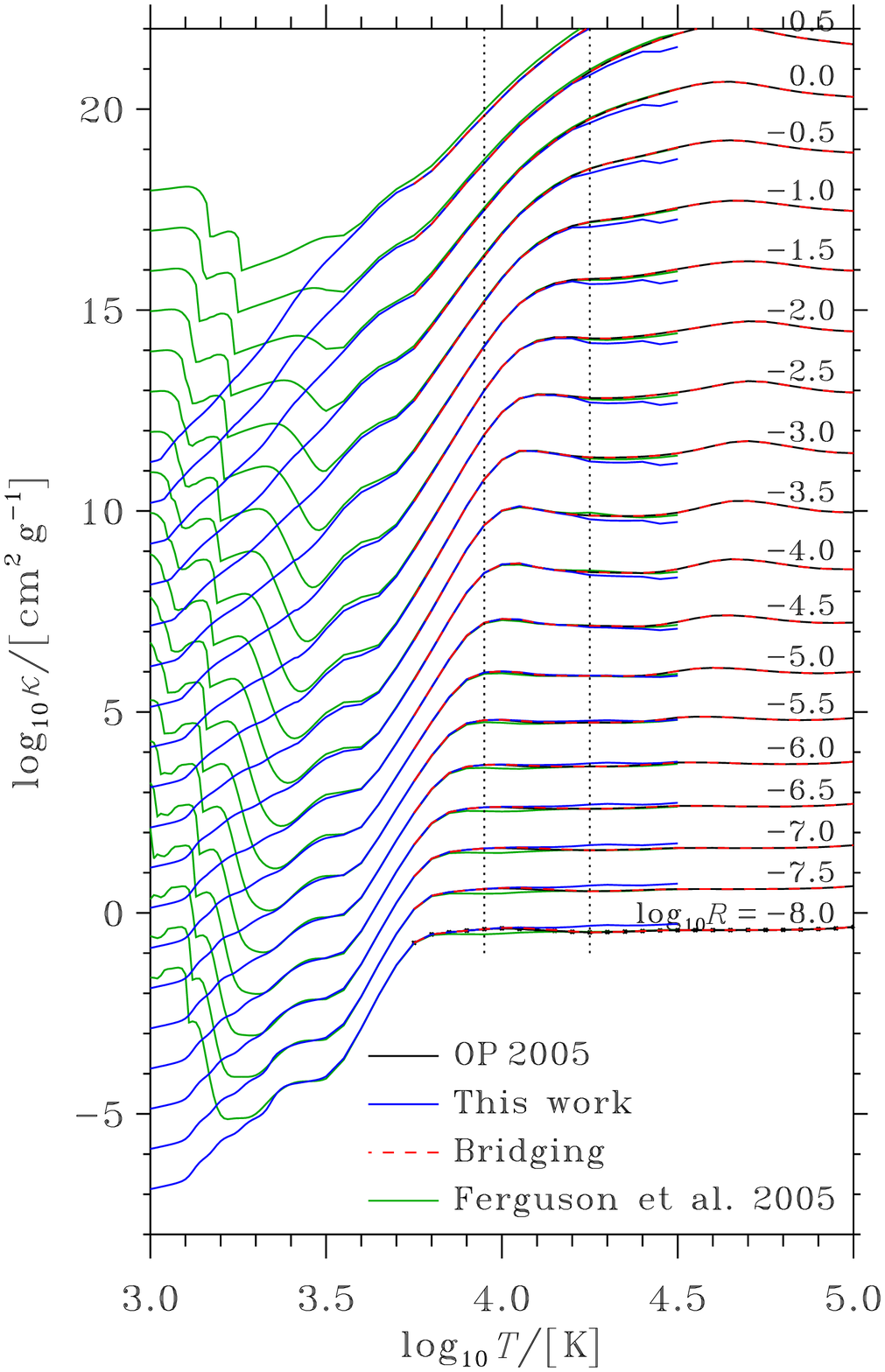}{12.4cm}
    {Opacities per mass as function of $\log T$. The 19 sets of curves are for
     $\log R=\log\varrho-3\stimes(\log T-6)$ as indicated, each off-set by 1.0
     from the previous, with no off-set for $\log R=-8$. The black curves show
     the interior OP \citep{OP05} opacities for $\log T\ge 3.75$. The blue
     curves show the present calculation of atmospheric opacities, compared
     with the \citet{ferguson:LowTOpac2005} opacities in green. The atmospheric
     are bridged with the interior opacities between the two vertical dotted
     lines.}

We have generated Rosseland opacity tables with these absorption sources, for
a rectangular grid of $(X,Z)$ compositions with eight values of $X\in[0; 0.9]$
and 13 values of $Z\in[0; 0.1]$. We have not merely scaled the contribution of
each opacity source by the change in abundance (i.e., assuming the EOS to 
be linear in composition), but have rather re-computed the EOS for each of the
104 sets of compositions to use as a basis for the summation of opacities.

These tables have in turn been assembled into
one global opacity table that can be interpolated in density, $\log\varrho$,
temperature, $\log T$ and composition, $X, Y$. These atmospheric, low-temperature
tables ($\log T\in[3; 4.5]$) have furthermore been merged differentiably with
interior opacities from the Opacity Project \citep[OP,][]{OP05}.
The combined opacities can be accessed by stellar structure codes through the
interpolation package, {\tt OPINT}, by \citet{houdek:gh_int.v6}, currently at
version 11.

We show our atmospheric opacities in Fig.\ \ref{AG89Fe.70.0200.tron} together
with the interior OP opacities. We note that the two sets of opacities show
good agreement in the temperature range where we bridge between the two. In
Fig.\ \ref{AG89Fe.70.0200.tron} we also compare with the atmospheric opacities
of \citet{ferguson:LowTOpac2005}, which also include water molecules
(contributing for $\log T\la 3.5$) and dust particles (the two bumps for
$\log T\la 3.2$).
None of our simulations enter the dusty regime, but our coolest simulations
would be affected by absorption by water.
%
\section{$T(\tau)$-Relations of the Simulations}
\label{simT-tau}

\begin{figure*}
    \centerline{\psfig{figure=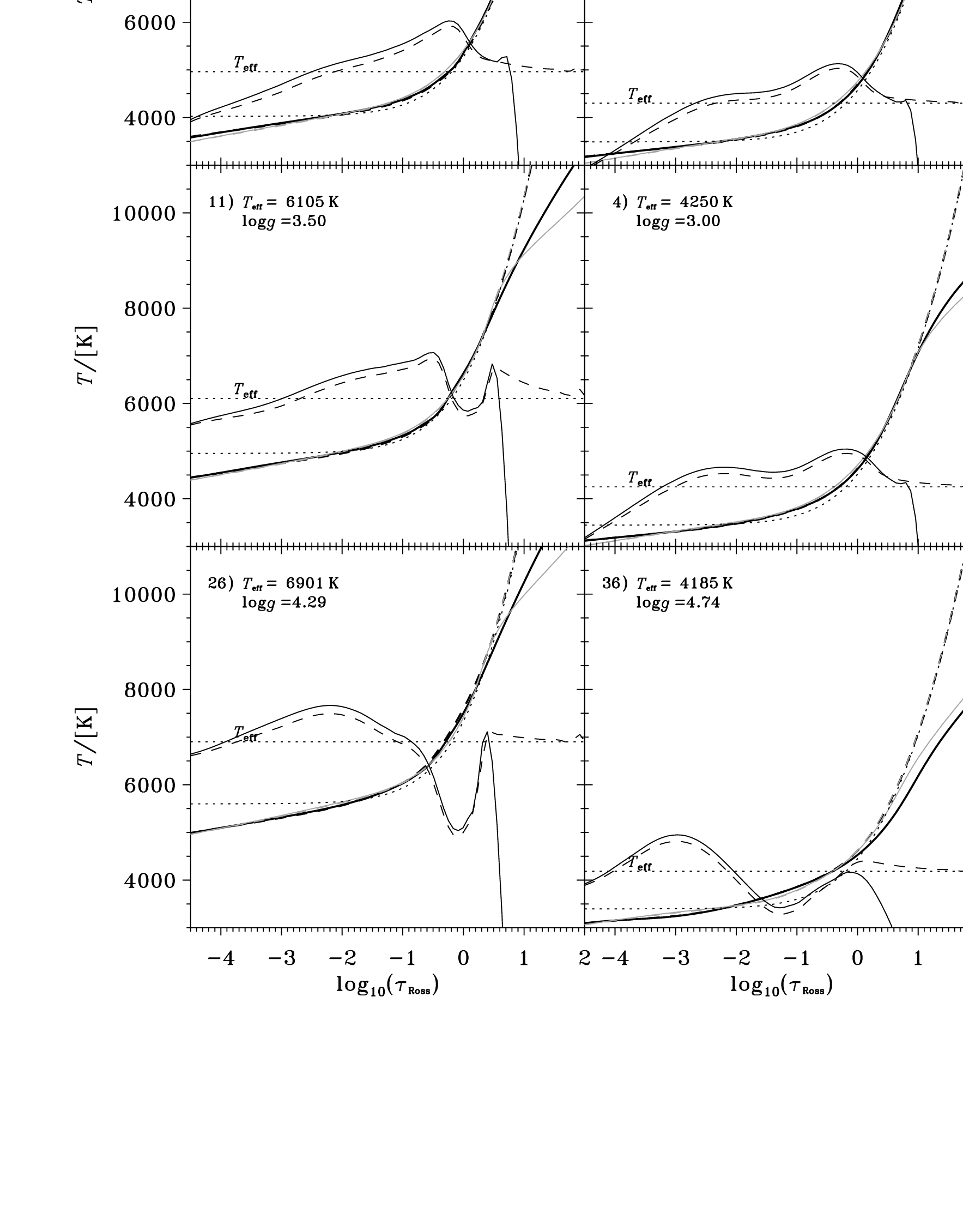,width=16.6cm}}
    \caption{\label{ttau_scan}
    Comparison of {\Ttau}s for six of the simulations with corresponding 
    1D atmosphere models. Solid lines show full {\Ttau}s,
    including the first-order convective effect, and 
    dashed lines show radiative {\Ttau}s without this first-order convective
    effect. The black thick lines show results for the 3D simulations, the          thin grey lines show 1D atmospheres
    interpolated in the MARCS \citep{MARCS-2} grid. The solution to the
    grey atmosphere is shown with dotted lines.
    The horizontal dotted lines shows $T_{\rm eff}$.
    The differences between the 1D MARCS models and the 3D simulations are shown
    around the $T=T_{\rm eff}$ line with thin black lines (dashed for the
    radiative {\Ttau}s), and exaggerated by a factor of 10. The atmospheric
    parameters and index in Table \ref{simlist} are given in each panel.}
\end{figure*}
The {\Ttau}s of six of the simulations (Nos.\ 2, 3, 11, 4, 26 and 36 of Table
\ref{simlist}) are compared with 1D MARCS stellar atmospheres in
Fig.~\ref{ttau_scan}. The thick solid lines show the partially convective {\Ttau}s of the
simulations, and the thick dashed lines show the
radiative {\Ttau}s, where the
first-order convective effect has been removed.

%
The light grey lines
show the MARCS 1D models by \citet{MARCS-2},
with the dashed grey lines showing the radiative {\Ttau}s, as calculated from
Eqs.~(\ref{qqrad}) and (\ref{fdBRoss}).
The 1D MARCS models are interpolated to the same atmospheric parameters as the
simulations. The corresponding 1D ATLAS9 models are very similar to the MARCS
models, with only minor differences below the photosphere due to different choices for
the parameters of the standard MLT formulation \citep{boehm:mlt} of convection
[see \citet[App.~A]{ludwig:alfa-cal} for a comparison of the
form factors of the various flavours of MLT].

The plots of Fig.~\ref{ttau_scan} are ordered with $T_{\rm eff}$ decreasing
to the right, and $\log g$ decreasing towards the top. This ordering also 
means that the most efficient convection occurs in the right-side plots, and
the least efficient on the left side. Low efficiency of convection means that it
takes larger velocities to transport the required flux, resulting in the
turbulent pressure contributing more to the hydrostatic equilibrium
(sub-photospheric maxima of the $p_{\rm turb}/p_{\rm tot}$-ratios of 27\%
for our warmest dwarf, No.\ 26, vs.~3.9\% for our coolest dwarf, No.\ 36, of table \ref{simlist}),
expanding the atmosphere \citep{trampedach:3Datmgrid}. It is also
accompanied by larger temperature fluctuations, not the least due to the large
temperature sensitivity of the opacity. The higher opacities in the warm
upflows mean that they stay adiabatic until (spatially) very close to the
photosphere. These are the bright granules. In the cooler downflows, the
photosphere lies much deeper with the overturning plasma radiating away energy
all the way. The net effect is higher temperatures below the photosphere,
compared with a corresponding 1D model. This effect is also known as convective
back-warming \citep{trampedach:SOHO12}, and expands the atmosphere by about as
much as the turbulent pressure (as higher temperatures cause larger pressure
scale-heights).

From Fig.~\ref{ttau_scan} we clearly see the first-order effect of convection
(due to a non-constant radiative flux), as the differences between
solid and dashed lines. This shows the onset of convection and
the sets of lines converge above the convection zone.

The difference between the 1D atmospheres (dotted and short dashed lines)
and the grey atmospheres (solid grey lines) in the photosphere and above,
shows the effect of line-blanketing. The line-absorption evidently affects
the cooler atmospheres the most, although they all show the signature of
strong lines, by having a temperature gradient even at the uppermost point
of the model.

The radiative {\Ttau}s of the simulations (dashed lines) converge to the grey
atmospheres (grey lines) with depth, as expected. In the photosphere and
above, the deviations show the effects of lines and higher-order convective
effects. In this region the differences between the 1D models and the 3D
simulations mainly show the higher-order convective effects. This comparison
can be most cleanly carried out with the ATLAS9 models since we have employed
the same line opacity data for our 3D simulations. The higher-order
convective effects are clearly significant and depend in a non-trivial way
on the atmospheric parameters.

\section{The Solar Case}
\label{solarT-tau}

Simulation No.\ 30, presented in Table~\ref{simlist}, is constructed to
correspond to the Sun.
\mfigur{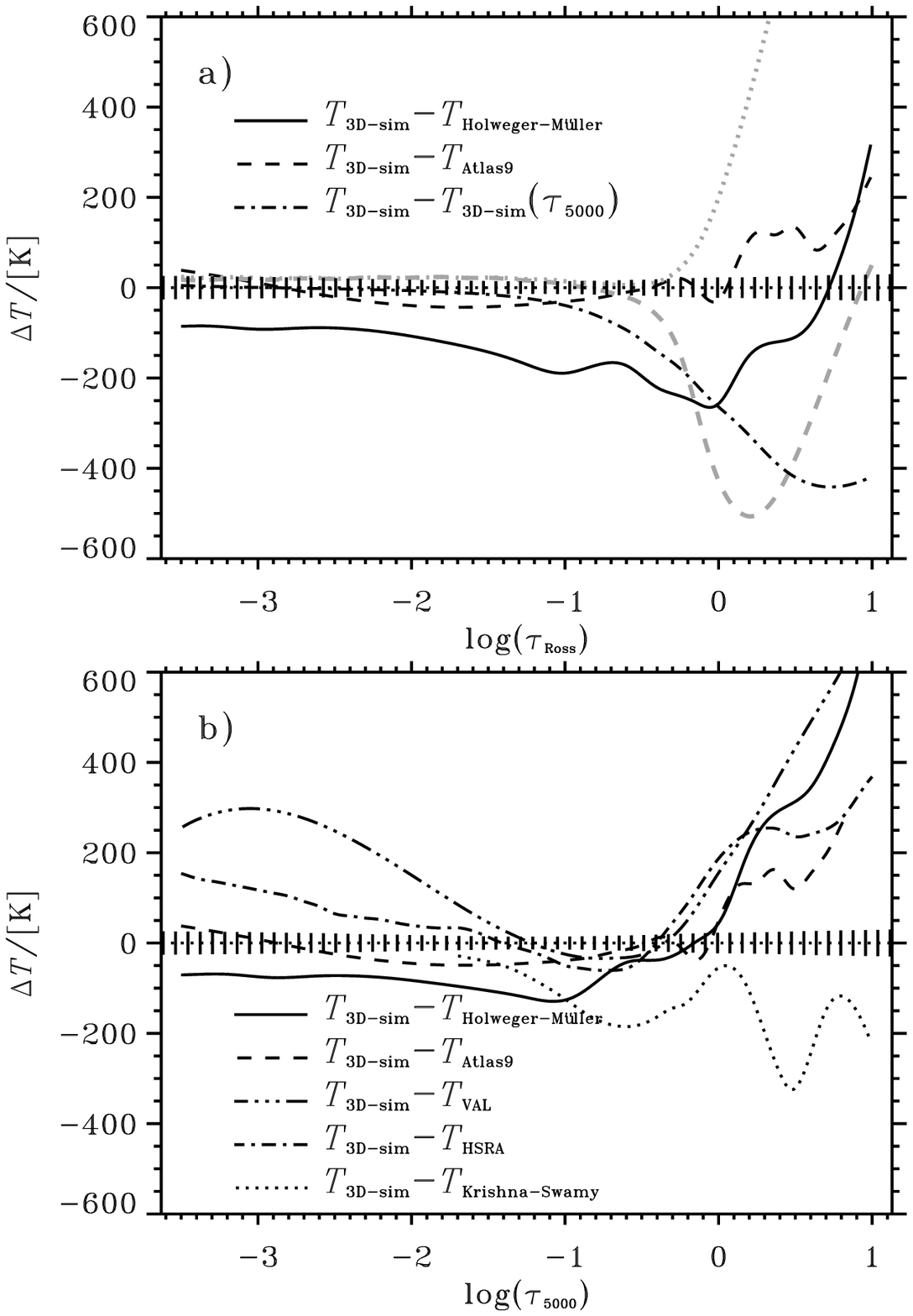}{12.4cm}
    {Comparison of the full $\tau$-averaged temperature structure,
     $\langle T\rangle_\tau$, from the solar simulation, with some often
     used solar {\Ttau}s. {\bf a)} The difference in Rosseland
     {\Ttau}s between the simulation and the
     semi-empirical model by {\protect\citet[solid line]{holweger-muller}},
     and an ATLAS9 atmosphere model
     {\protect\citep[dashed line]{atlas9,castelli:atlas9-conv}}.
     Here we also show the
     difference between the two $\tau$-scales (dot-dashed line) as
     $T(\tau_{\rm Ross}=x)-T(\tau_{\rm 5000}=x)$, as well as comparisons
     with the averaging methods of Fig.\ \ref{mT-tau} in grey: dotted for case
     b) $\langle T\rangle_{\rm L}(\langle\tau_{\rm Ross}\rangle_{\rm L})$ and
     dashed for case c) $\langle T\rangle_{\rm L}(\tau_{\rm Ross}
     (\langle\varrho\rangle_{\rm L},\langle T\rangle_{\rm L}))$.
     {\bf b)} Differences, on a
     $\tau_{5000}$-scale, between the simulation and an ATLAS9 model,
     and four semi-empirical atmosphere models: The model by
     {\protect\citet{holweger-muller}}, the classical fit by
     {\protect\citet{krishna-swamy:lines-K-dwarfs}},
     the HSRA model {\protect\citep{HSRA}} and the average, quiet-Sun, VAL model
     {\protect\citep[see text]{VAL-III}}.}
It has an effective temperature of $T_{\rm eff}=5\,774\pm 17$\,K, 
and a (nominal) surface gravitational acceleration of 27395.9\,cm\,s$^{-2}$.

We compare the resulting {\Ttau}s with various 1D models 
in Fig.~\ref{sunT-taudiff},
using both the optical depth $\tau_{\rm Ross}$ defined with
the Rosseland mean opacity $\kappa_{\rm Ross}$
and the monochromatic optical depth $\tau_{5000}$ corresponding to
the opacity at 5000\,{\AA}. This plot compares the full {\Ttau}s including
convection,
employing flavours of MLT in the 1D theoretical models.
Panel a) shows temperature differences on the
$\tau_{\rm Ross}$-scale and panel b) is for the $\tau_{5000}$-scale.
The vertically hatched area around the zero-line (in both panels) shows
the temporal RMS-scatter of the {\Ttau} of the simulation, indicating
which differences are statistically significant.
Panel a) also shows the difference between the temperature measured
on the two $\tau$ scales for the simulation (dot-dashed curve),
the sign of which
means that we can see deeper into the Sun at 5000\,{\AA} than on (a Rosseland)
average, as a consequence of the fact that the Rosseland opacity is larger than the 5000\,{\AA}
opacity.

The past two decades of work on compiling and computing line data for atoms
and molecules has added a lot of line opacity in the UV, which has increased
the Rosseland opacity with respect to the 5000\,{\AA} opacity
\citep{kur:missolar,kurucz:newATLAS9,trampedach:thesis,MARCS-2}. This in turn has
increased the difference between $\tau_{\rm Ross}$ and $\tau_{5000}$, which
is clearly expressed in the 200\,K difference at $\tau\simeq\frac{2}{3}$.
This difference is twice as large as the corresponding
differences amongst modern atmosphere models ({\cf} Fig.~\ref{sunT-taudiff}b),
so it is no longer justified to assume the two $\tau$ scales to
be equal.
For stellar structure calculations, it has been common practise to use a
$T$-$\tau_{5000}$ relation
combined with a Rosseland opacity. With the current opacities and
today's demand for accuracy, this no longer seems a valid approximation
and we recommend against using it.

The semi-empirical \citet{holweger-muller} model 
(solid line in Fig.~\ref{sunT-taudiff}) is given on both the 5000\,{\AA} and the
Rosseland $\tau$-scales. We have used their original data and have made no attempt to update the Rosseland
scale with modern opacities. Above the photosphere the \citet{holweger-muller}
model is about 100\,K warmer than the simulation,
on both $\tau$ scales. For this model the $T$ difference between
the two $\tau$ scales is less than 60\,K, so in layers deeper than
$\tau\simeq 0.5$
the behaviour in the two panels is dominated by the 
$(\tau_{\rm Ross}-\tau_{5000})$ difference {\em for the simulation}.
In the photosphere
the Holweger-M\"uller model is hotter than the simulation by
up to 300\,K on the $\tau_{\rm Ross}$-scale, but on the $\tau_{5000}$-scale
they are very similar. On both $\tau$-scales the Holweger-M\"uller model
becomes increasingly cooler with depth, compared with the 3D solar simulation, indicating
a faster transition to convective transport of the whole flux.

For the theoretical ATLAS9 atmosphere models \citep{atlas9}, both
monochromatic
and Rosseland {\Ttau}s are available, illustrated as the dashed line in both
panels of Fig.~\ref{sunT-taudiff}.
The peculiar wiggles in these curves are features from the ATLAS9 model.
Using the ``overshoot''-option, later disfavoured by
\citet{castelli:atlas9-conv}, these wiggles combine to a larger but smoother
dip, compared with the simulation. The rather close agreement in the radiative
part of the atmosphere is expected, since the same line opacities were used,
and since the convective fluctuations have only a small effect on the averaged
{\Ttau} above the convection zone
({\ie}, all averaging methods give the same results, {\cf} Fig.~\ref{mT-tau}).
This is not necessarily the case for other stars where there will be a different
balance between the expansion cooling of the overshooting flows and the
radiative heating/cooling by spectral lines
\citep[see Sect.\ \ref{parvar},][]{asplund:low-Z-Li,remo:3DRedGiantAbunds}---a
balance that does not exist in 1D models due to a lack of physical models of
overshooting.

The two other semi-empirical atmosphere models presented in
Fig.~\ref{sunT-taudiff},
the component averaged VAL model for the quiet Sun \citep{VAL-III} and the
Harvard-Smithsonian solar
reference atmosphere (HSRA) \citep{HSRA}, differ significantly
and essentially
in the same way from the simulation results.
We used an average of the VAL model over its six components, with quiet-Sun
weights as specified in their Table 7. This structure is close to the often
used C component, but is warmer by 2\,K around the photosphere, rising to
27\,K at $\log\tau_{5000} = -2.25$. Most of the differences between VAL-C and
the average VAL model, stay below the RMS-fluctuations of our solar simulation.

High in the atmosphere, the situation is a bit less clear. This plot covers
up to 0.5\,Mm above the photosphere, although the simulation reaches up to
0.8\,Mm above. The solar temperature minimum occurs around 0.5\,Mm before the
sudden rise to chromospheric temperatures at an average height of about 2\,Mm.
In the VAL\,C model
the temperature minimum occurs at $\log\tau_{5000}\simeq -3.5$, but there is
no sign of temperature minima in any of our simulations.
%
%
This is not surprising, considering the lack of non-LTE effects or magnetic
fields in our simulations, and their implications in chromospheric and coronal
heating \citep[e.g.,][]{gudiksen:CoronHeat,vanballegooijen:CoronAlfvenHeat}.

The last {\Ttau} presented in Fig.~\ref{sunT-taudiff} is the one by
\citet[dotted line, only defined for $\tau=0.02$--10]{krishna-swamy:lines-K-dwarfs}, which is still
used as an upper boundary in some stellar model codes \citep[{\eg},][]{straniero:FRANEC3.1,chaboyer:ZdiffInLowZstars,pietrinferni:BastiI,demarque:YREC}.
It is interesting to
note that the behaviour below the photosphere is opposite that of the more
modern {\Ttau}s.

At intermediate optical depths, from $\log\tau \simeq -1$ down to
$\langle T\rangle = T_{\rm eff}$, the agreement between the simulation and both
theoretical and semi-empirical atmospheres is rather good, as expected.
Differences are also smaller than the difference between the 5000\,{\AA} and the
Rosseland {\Ttau} from the simulation.

The agreement with semi-empirical models in this region is encouraging, as
convective fluctuations are smaller here, damped by radiative losses. It is, however, not in itself a
validation of the convection simulations. Such a validation has to be
performed much closer to the actual observations, in order to
avoid all the theoretical biases that go into, {\eg}, semi-empirical models.
Semi-empirical atmospheres have too often been assumed equivalent to
observations, and we argue that the considerable differences between the VAL,
HSRA and Holweger-M\"uller models speak against that practise.
The actual limb darkening, spectral energy distribution and line profiles are
the benchmarks 
with which all atmosphere models should be compared.
Solar and stellar 3D simulations
presented here perform very well against observational diagnostics, including detailed line profiles
\citep{asplund:solar-Fe-shapes,prieto:Procyon-conv-Fe}, and the solar
continuum limb darkening
\citep{tiago:suneHlinesLimbdark1a,tiago:suneHlinesLimbdark}.

As convection becomes the dominant mode of energy transport, the {\Ttau}s in
Fig.~\ref{sunT-taudiff} diverge,
with the various theoretical 1D models, sharing the MLT formu\-lation of convection, all
differing from the simulation in more or less the same manner.
The semi-empirical models do not account for convection.
Below $\langle T\rangle = T_{\rm eff}$, 3D effects
become important and the turbulent pressure contributes up to 14\% of the
total pressure, rendering the simulation the better choice for an atmosphere
model.

\section{Variation with Stellar Atmospheric Parameters}
\label{parvar}

Having evaluated {\Ttau}s for our grid of simulations, we can now proceed to
give an overview of the behaviour with atmospheric parameters, $T_{\rm eff}$
and $\log g$.

The $q$ surfaces as function of stellar parameters, are shown at
three optical depths in Fig.~\ref{simgriddat_scan}. The edge towards the
reader is populated by main-sequence stars and the far right corner, by red
giants. The offset between the three surfaces is the real difference between
the Hopf functions at the three $\tau_{\rm Ross}$-levels.
The bottom level in the plot (highest point included in the
atmosphere) shows fairly little variation.
At the mid-level, there is an increase of $q$
towards the cool dwarfs and a slow increase towards the giants.
In the optical deep part, $q\rightarrow q_\infty$, is dominated by
a broad bump at medium-temperature dwarfs that extends towards the cool giants,
and an upturn on the main-sequence for both hot and cool dwarfs. The location of
the solar simulation is indicated with $\odot$ and lies near the maximum of the
bump in $q_\infty$, which means $q$ changes little in the immediate
neighbourhood of the Sun.
\mfigur{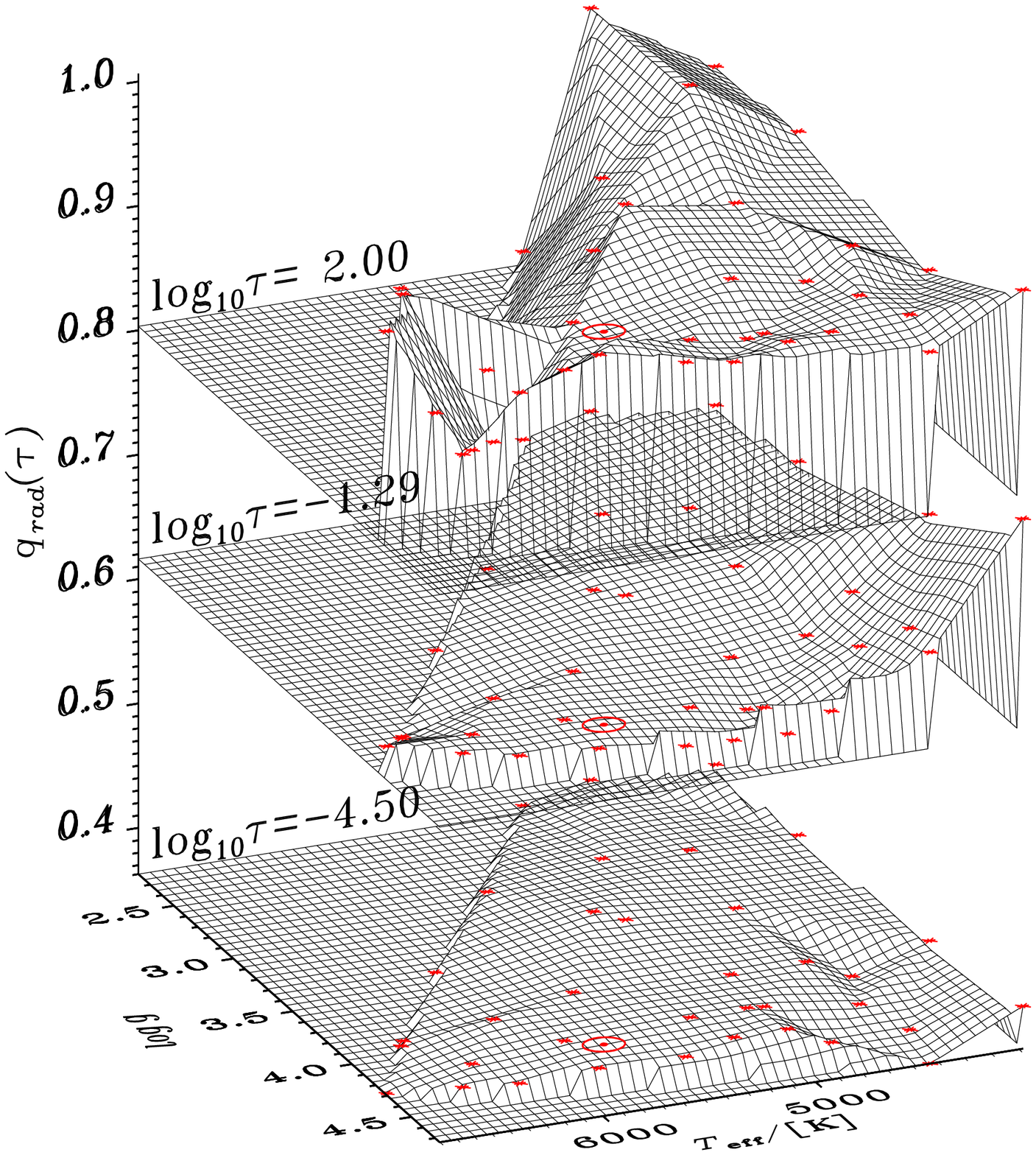}{10.0cm}
    {The dependence of the radiative Hopf function, $q$ as function
     of optical depth and atmospheric parameters, $T_{\rm eff}$ and $\log g$.
     The location of the simulations
     are indicated with red asterisks, and the solar simulation is indicated
     by a red $\odot$.}

\mfigur{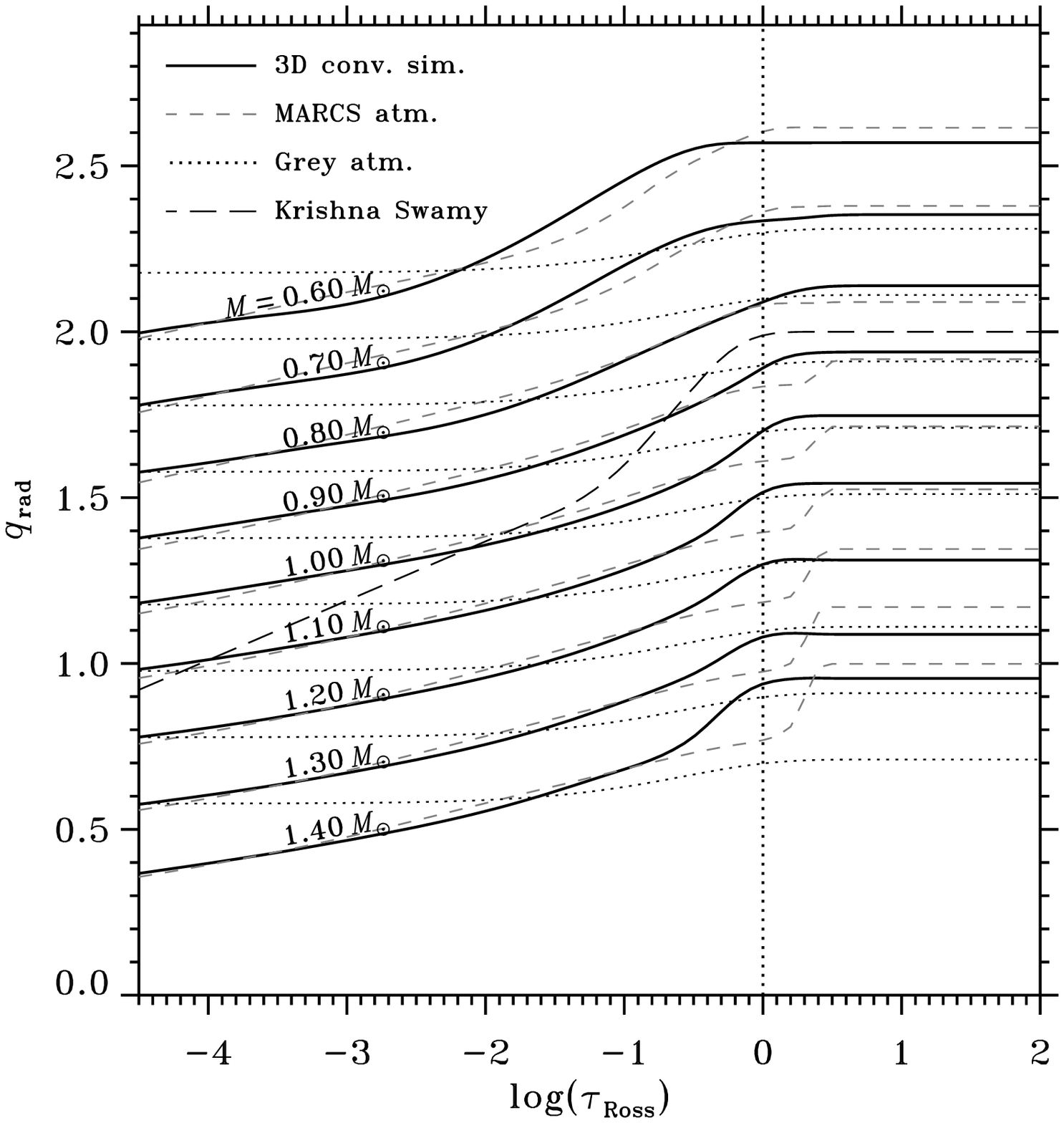}{9.3cm}
    {The change of the radiative Hopf function $q$ with stellar
     mass, on the zero-age
     main sequence. The mass is indicated to the right, and each set of curves is
     offset by 0.2 with respect to the next mass, with the $M=1.40\,M_\odot$ curve
     having no offset. The solid lines show the simulations interpolated to
     the parameters of the zero-age main sequence, 
     and the dashed grey curves show the corresponding 1D MARCS models
     \citep{MARCS-2}.
     The dotted curves indicate the grey atmosphere, crossing their
     corresponding line-blanketed atmospheres between $\log\tau_{\rm Ross}=-2$
     and $-1.5$. The long-dashed curve
     shows the often applied {\protect\citet{krishna-swamy:lines-K-dwarfs}} relation,
     offset by the same amount as the $1.00\,M_\odot$ curves.
     The vertical dotted line indicates optical depth unity.}
In Fig.~\ref{ZAMSTtau} we present the variation of $q$ with
stellar mass on the zero-age main sequence (ZAMS). The atmospheric
parameters for the ZAMS were derived from the stellar evolution models computed
with the MESA-code \citep{paxton:MESA}, using MLT $\alpha$ and initial helium
abundance $Y_0$ calibrated to the present Sun.
The radiative Hopf functions were interpolated linearly on the Thiessen
triangles of the irregular grid. For comparison we also plot
$q$ derived from 1D MARCS atmospheres (grey dashed lines in
Fig.~\ref{ZAMSTtau}), through Eqs.~(\ref{qqrad}) and (\ref{fdBRoss}).
The simulations have a smooth transition to the asymptotic value in the
diffusion approximation, $q_\infty$,
contrary to the behaviour of the $M>0.8\,M_\odot$ 1D MARCS models, which exhibit
a deeper and rather abrupt transition to $q_\infty$ and the medium-mass models
also display a marked plateau around the photosphere. 
We suspected second-order effects from the onset of MLT convection, but this feature in $q_{\rm rad}(\tau)$ does not seem
to be correlated with the $F_{\rm conv}/F_{\rm tot}$-ratio of the models.
The optically deep values, $q_\infty$, are lower for the 3D simulations than
their 1D counterparts for high and low masses and opposite around 1\,$M_\odot$.
The 3D {\Ttau}s above the photosphere are generally seen to have a larger
range of behaviour compared with those of the 1D models; The 3D case is
shallower in the high atmosphere and steeper at intermediate $\tau$.
The low-mass simulations converge to $q_\infty$ strikingly high in the atmosphere.
This is not caused by interpolation between the simulations, as the
parameters of the zero-age, 0.6\,$M_\odot$ star are very close to 
simulation No.\ 36 of Table \ref{simlist}, which displays a similar behaviour.

Figure~\ref{GRAVTtau} illustrates the gravity dependence of the {\Ttau}s.
Going to lower gravity, the {\Ttau} gets steeper in the
photosphere and shallower just above, developing a kink right in the
photosphere of the lowest gravity case. This feature is also not a result of
the interpolation between the simulations, since simulation No.\ 2 is very close to
the $\log g=2.25$ case plotted, and exhibits the same kink.
The $q$s from the 1D models are generally steeper
in the high atmosphere, and show very little overall variation with gravity,
compared with the 3D simulations.

The opacities in the ATLAS9 models and
the 3D simulations are nearly identical, and \citet{MARCS-2} found that the
\mfigur{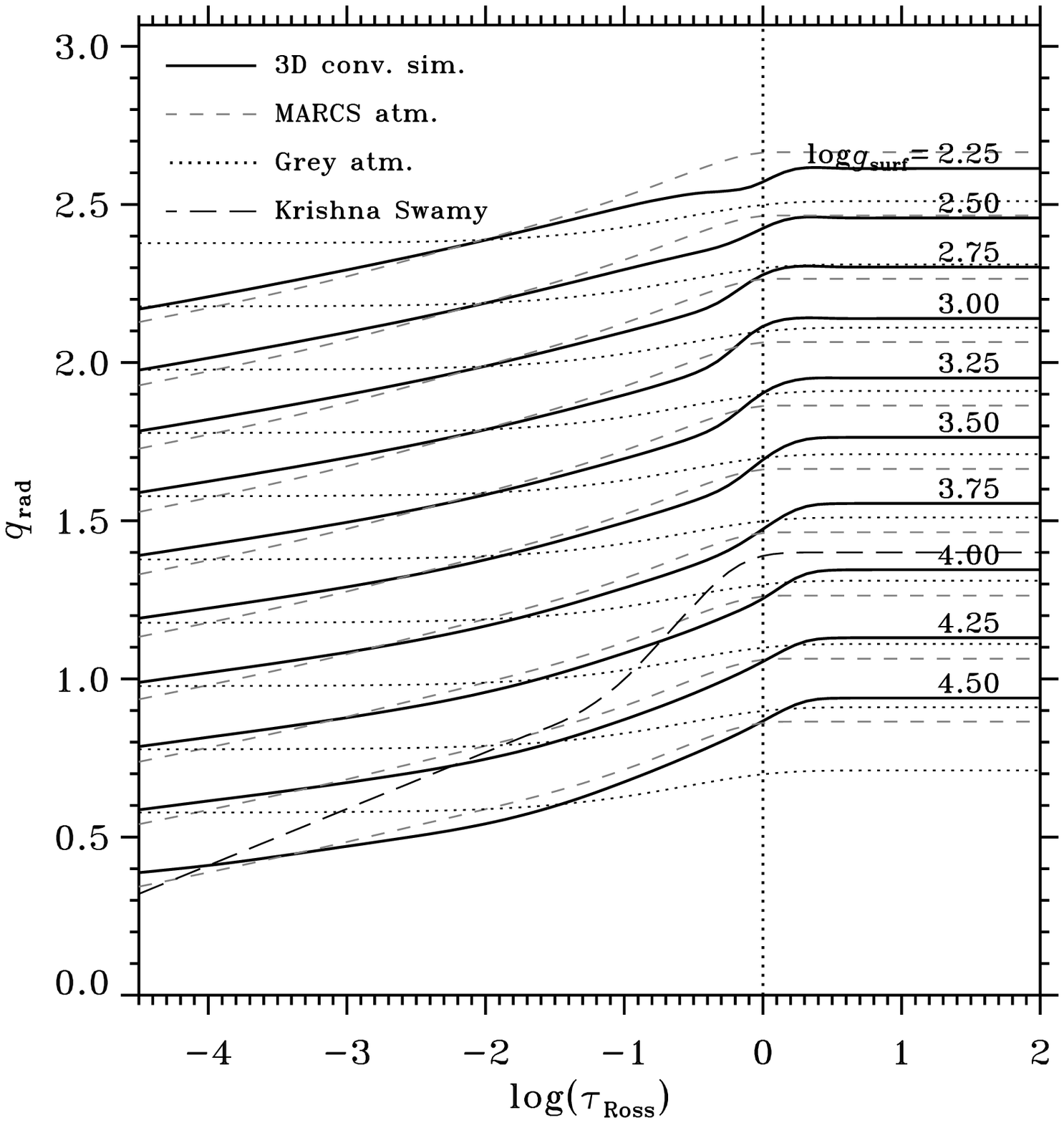}{9.3cm}
    {As Fig.~\ref{ZAMSTtau}, but showing the change of $q$ with
     gravity, for a fixed $T_{\rm eff} = 5000$\,K. This $T_{\rm eff}$ is chosen
     for the grid of simulations having the largest extent in gravity here.
     The offset between each set of curves is 0.2, and the $\log g=4.5$
     curves have no offset. The {\protect\citet{krishna-swamy:lines-K-dwarfs}}
     atmosphere has no offset either.}
differences between MARCS and ATLAS9 models in this region amount to less
than 20\,K throughout the atmospheres, $-5\le \log\tau \le 2$.
This 
means that the differences we see here, between 1D atmospheres and 3D
convection simulations, must be mainly due to 
the more realistic treatment of convection in our simulations.

One of the major differences between hydrodynamic simulations and the MLT
formulation is the presence of overshooting from the top of the convection
zone and into the stably stratified part of the atmosphere. As a result, we
see significant velocity fields throughout the atmospheres of our simulations.
As pointed out by \citet{asplund:low-Z-Li}, 
the radiative heating and cooling therefore have competition from
the expansion cooling of rising plasma, expanding due to the large density
gradient in the atmosphere. The adiabatic stratification is typically more
than 1000\,K cooler than the radiative-equilibrium solution. A stratification
in-between these two extremes will therefore experience radiative heating
and expansion cooling, and obtain equilibrium at a lower temperature than
the purely radiative atmosphere.
This convective cooling from overshooting is one of the higher-order convective
effects visible in Figs.~\ref{ZAMSTtau} and \ref{GRAVTtau}.
And the process occurs outside
the convection zone, in a region where convection carries no flux --- or
rather, the convective flux is less than a percent of the total and it is
negative. This effect is rather small for solar metallicity stars, as the large
opacity will establish an equilibrium close to the radiative one.
For metal-poor stars the effect can be of order 
$10^3$\,K \citep{asplund:low-Z-Li,remo:3DRedGiantAbunds}.

In Figs.~\ref{ZAMSTtau} and \ref{GRAVTtau} we also show the
grey atmosphere \citep{king:GrayAtm,mihalas:stel-atm} with dotted lines.
It is obvious that the grey approximation is insufficient in all cases.

\section{Conclusion}
\label{conclusion}

We have confirmed that the use of {\Ttau}s is indeed a good and consistent method
for incorporating
the effects of full radiative transfer in stellar structure computations -- even in the
non-grey case, as shown in Sect.~\ref{Ttaubasis}.
It is often remarked that {\Ttau} relations imply grey radiative transfer
since the are used with the grey Rosseland opacity. That is \emph{only} the
case, however, when the grey Hopf function is used. When a generalized Hopf
function is used, any physical effects can be included in the atmosphere, and
with the {\Ttau} and a Rosseland opacity the stratification can be
reconstituted without the need for detailed radiative transfer.
With our formulation the
{\Ttau}s can be based on any realistic atmosphere calculation, be it 1D, 3D or non-LTE, without the 
stellar structure code implementing them needing any information about such
complications. We also developed a more general formulation to deal with
{\Ttau}s in 3D, which reduces properly to the simpler 1D case.

Based on that we proceeded to compute {\Ttau}s for a number of 3D
simulations of radiation-coupled convective stellar atmospheres, the results of
which are displayed in Sects.~\ref{simT-tau} and \ref{solarT-tau}.
Comparisons with 1D models reveal differences of the order of
100\,K in the shape of the {\Ttau}s (see Fig.\ \ref{ttau_scan}). In
Figs.\ \ref{ZAMSTtau} and \ref{GRAVTtau} the differences are put in the context
of stellar evolution, by displaying cuts in the HR diagram. These differences
are again significant and smooth but not monotonic in atmospheric parameters, with the 3D simulations displaying
a larger range of behaviours---especially with varying gravity and fixed
$T_{\rm eff}$. The warmer 1D MARCS atmospheres show a curiously sharp
transition to the constant Hopf function of the diffusion approximation.

The simulations populate the $T_{\rm eff}/g$-plane densely enough for
interpolation between the simulations to be safe. Routines for
interpolating irregularly gridded data are publicly available
\citep{renka:triangulation} and makes for
straightforward implementation in stellar structure and evolution codes.

To separate, in stellar structure
models, the effects of convection from those of the radiative transition
in the photosphere, and to avoid unnecessary systematic effects,
we recommend that the {\Ttau}s be reduced to radiative equilibrium, as defined in
Sect.~\ref{Ttaubasis}, and that convection is reintroduced into the interior
structure model,
as described in Sect.~\ref{Ttau-impl}.

As the precision and scope of modern observations of stars steadily improve, and
as we find ourselves in the age of detailed asteroseismic inferences, higher demands are placed on
the modelling of stars. With improved understanding and treatment of the
interplay between radiation and convection it will be possible to isolate
other effects that so far have been shrouded in the uncertainty of the
atmospheric part of stellar models. With improved outer boundary conditions,
combined with the mixing-length calibration in Paper II, we can have more
confidence in predictions about the depth of convective envelopes. This,
in turn, will allow the study of other mixing processes, such as convective
overshoot at the base of the convection zone,
rotational and gravity-wave mixing, 
{\etc}, and comparisons with observations of chemical enrichment from
dredge-ups and the destruction of volatile elements such as Li and B.
\aatab*{tab:q}{Sample of the on-line table of Hopf functions as functions of $\log\tau$ (rows) and atmospheric parameters, $T_{\rm eff}$ and $\log g$ (columns)}{lrrrrrrrrc}
{  sim.\ no. &  \oh{18}    &  \oh{30}    &  \oh{28}    &  \oh{17}    & \oh{26}     &  \oh{~3}    &  \oh{16}    &  \oh{24}   &$\cdots$\\}{
$T_{\rm eff}$/[K]&6582.301 &    5774.501 &    4813.199 &    6137.301 &    6901.799 &    4301.201 &    5674.800 &    5926.601 &$\cdots$\\
$\log g$     &       3.966 &       4.438 &       4.500 &       4.040 &       4.292 &       2.420 &       3.943 &       4.295 &$\cdots$\\
$[$Fe/H]     &       0.000 &       0.000 &       0.000 &       0.000 &       0.000 &       0.000 &       0.000 &       0.000 &$\cdots$\\
 MLT alpha   &  1.64810097 &  1.76738799 &  1.76003206 &  1.69658005 &  1.67718101 &  1.76493704 &  1.75717497 &  1.75384402 &$\cdots$\\
 sig alpha   &  0.02925100 &  0.03049900 &  0.02174600 &  0.02526400 &  0.03825700 &  0.02123500 &  0.03448400 &  0.02877900 &$\cdots$\\
 -4.50000000 &  0.37156389 &  0.38330216 &  0.36864682 &  0.38466947 &  0.36400630 &  0.39851932 &  0.38638091 &  0.38371588 &$\cdots$\\
 -4.41975307 &  0.37620499 &  0.38797457 &  0.37335618 &  0.38895423 &  0.36898350 &  0.40410561 &  0.39158757 &  0.38808286 &$\cdots$\\
 -4.33950615 &  0.38091088 &  0.39281506 &  0.37816540 &  0.39337589 &  0.37398517 &  0.40972824 &  0.39694869 &  0.39258415 &$\cdots$\\
 -4.25925922 &  0.38565236 &  0.39779905 &  0.38305531 &  0.39795361 &  0.37900728 &  0.41537387 &  0.40240468 &  0.39723357 &$\cdots$\\
 -4.17901230 &  0.39043554 &  0.40290439 &  0.38791891 &  0.40268837 &  0.38405111 &  0.42102208 &  0.40789370 &  0.40201536 &$\cdots$\\
 -4.09876537 &  0.39529022 &  0.40812081 &  0.39283812 &  0.40756858 &  0.38909873 &  0.42666348 &  0.41341677 &  0.40693703 &$\cdots$\\
 -4.01851845 &  0.40023393 &  0.41343065 &  0.39790501 &  0.41258876 &  0.39417138 &  0.43230324 &  0.41897563 &  0.41198729 &$\cdots$\\
\oh{$\cdots$}&\oh{$\cdots$}&\oh{$\cdots$}&\oh{$\cdots$}&\oh{$\cdots$}&\oh{$\cdots$}&\oh{$\cdots$}&\oh{$\cdots$}&\oh{$\cdots$}&$\cdots$\\
}

\paragraph*{{\bf Data Retrieval:}} A file with the $q(\tau)$ data and
Fortran\,77 routines for reading and interpolating the data can be downloaded
from: \url{http://www...}. A few entries from the on-line file are shown 
in table \ref{tab:q}.
The data file contains both the radiative Hopf functions, $q(\tau_{\rm Ross})$
and the calibrated mixing-length parameter, $\alpha$, as found in Paper II,
as functions of atmospheric parameters, $T_{\rm eff}$ and $\log g$. The URL
also contains the routines necessary for setting up and interpolating in the
triangulation of the irregular grid of simulations \citep{renka:triangulation}. 
Finally, we also supply a simple user-level function to include in
stellar structure codes, which does not require any knowledge of the data or
the details of the triangulation.

The {\tt OPINT} opacity interpolation package can be downloaded from
\url{http://phys.au.dk/~hg62/OPINT}, together with the atmospheric
opacities from our calculation, merged with interior OP opacities
(cf. Sect.\ \ref{sect:opac}).

\section*{Acknowledgements}

The helpful comments and suggestions by the anonymous referee have improved
the paper and are much appreciated.
We are grateful to R.\,L.\,Kurucz for providing access to his database
of opacity distribution functions, and we are grateful to
W.\,D{\"a}ppen for access to the code and data tables for the MHD equation
of state. RT acknowledges funding from the Australian Research Council (grants
DP\,0342613 and DP\,0558836) and NASA grants NNX08AI57G and NNX11AJ36G.
RFS acknowledges NSF grant AGS-1141921 and NASA grant and NNX12AH49G.
Funding for the Stellar Astrophysics Centre is provided by The Danish National Research Foundation (Grant DNRF106).
The research is supported by the ASTERISK project (ASTERoseismic Investigations with SONG and Kepler) funded by the European Research Council (Grant agreement No.: 267864).
The work of MA has been supported through a Laureate Fellowship from the 
Australian Research Council (FL110100012).
The simulations were
run at the Australian Partnership for Advanced Computations (APAC).
This research has made extensive use of NASA's Astrophysics Data System.


\bsp

\label{lastpage}

\end{document}